\documentclass[12pt,a4paper]{article}
\pdfoutput=1
\usepackage{jcappub}

\usepackage{graphicx}
\usepackage{amsmath}
\usepackage{amssymb}
\usepackage{hyperref}

\usepackage{color}

\def\bea{\begin{eqnarray}}
\def\eea{\end{eqnarray}}
\def\ba{\begin{eqnarray}}
\def\ea{\end{eqnarray}}
\def\be{\begin{equation}}
\def\ee{\end{equation}}
 
\newcount\hour \newcount\minute
\hour=\time \divide \hour by 60
\minute=\time
\title{Halo-independent tests of dark matter annual modulation signals}

\def\kth{Department of Theoretical Physics,
School of Engineering Sciences, KTH Royal Institute of Technology,
AlbaNova University Center, 106 91 Stockholm, Sweden}

\author[a]{\textbf{Juan Herrero-Garcia}\vspace*{0mm}}
\affiliation[a]{\kth}

\emailAdd{juhg@kth.se}

\abstract{
New halo-independent lower bounds on the product of the dark matter--nucleon scattering
cross section and the local dark matter density that are valid for annual modulations of dark matter direct detection signals are derived. They are obtained by making use of halo-independent bounds based on an expansion of the rate on the Earth's velocity that were derived in previous works. In combination with astrophysical measurements of the local energy density, an observed annual modulation implies a lower bound on the cross section that is independent of the velocity distribution and that must be fulfilled by any particle physics model. In order to illustrate the power of the bounds we apply them to DAMA/LIBRA data and obtain quite strong results when compared to the standard halo model predictions. We also extend the bounds to the case of multi-target detectors.}


\begin{document}
\maketitle

\flushbottom

\section{Introduction} 

In order to analyze a direct detection (DD) signal \cite{Goodman:1984dc, Drukier:1986tm, Freese:1987wu} it is typically assumed that the dark matter (DM) has a Maxwellian distribution of velocities truncated at the galactic escape velocity. This is known as the Standard Halo Model (SHM). However, $N$-body simulations indicate a more
complicated DM halo~\cite{Vogelsberger:2008qb, Kuhlen:2009vh, Kuhlen:2012fz}. The dependence of the WIMP signals on the velocity distribution has been studied in several works, see for instance refs.~\cite{Fornengo:2003fm, Green:2003yh,Fairbairn:2008gz, McCabe:2010zh, Green:2010gw, Freese:2012xd, Bozorgnia:2013pua}. In particular, the assumed velocity-distribution is crucial for the interpretation of annual modulation signals~\cite{Drukier:1986tm, Freese:1987wu}, with different unvirialized components such as tidal streams~\cite{Savage:2006qr, Gelmini:2000dm, Chang:2008xa, Natarajan:2011gz} or debris flows~\cite{Kuhlen:2012fz} modifying the results drastically~\cite{Freese:2012xd}. Due to these large uncertainties, halo-independent methods have been developed to compare different DD signals \cite{Drees:2007hr, Drees:2008bv, Fox:2010bz, Fox:2010bu, McCabe:2011sr, Frandsen:2011gi,
  HerreroGarcia:2011aa, HerreroGarcia:2012fu, DelNobile:2013cta,
  DelNobile:2013cva, Bozorgnia:2013hsa, Cherry:2014wia, Fox:2014kua,
  Feldstein:2014gza, Feldstein:2014ufa, Bozorgnia:2014gsa,
  Anderson:2015xaa,Scopel:2014kba,Kavanagh:2012nr,Kavanagh:2013wba}. In ref.~\cite{Blennow:2015oea} they have been extended in order to relate a positive signal in DD with the neutrino rate from DM annihilations in the Sun, while in ref.~\cite{Ferrer:2015bta} a halo-independent upper limit on the scattering cross section was obtained by combining upper limits from DD and neutrino rates from the Sun. 
  
Recently, in ref.~\cite{Blennow:2015gta} a new framework has been established that allows to compare the constant rate of a DD signal with local energy density measurements, LHC limits, the relic abundance and indirect detection constraints (see also ref.~\cite{Ferrer:2015bta} for another method to derive a lower bound on the cross section). In this work we show how the observation of an annual modulation signal in a DD detection experiment can also be used to place a lower bound on
the product of the local DM density and the scattering
cross section that is independent of the DM velocity
distribution, in a similar fashion as was done in ref.~\cite{Blennow:2015gta} for constant rates. For this purpose, we use halo-independent bounds on the annual modulation based on an expansion of the rate on the Earth's velocity~\cite{HerreroGarcia:2011aa, HerreroGarcia:2012fu}. In order to illustrate their use we perform a numerical analysis to DAMA/LIBRA data, which we will denote by DAMA in the following. As we will see, in combination with measurements of the local energy density we obtain quite strong lower bounds on the cross section, that must be fulfilled by any particle physics model in order for the signal to be consistent with DM.

The paper is structured as follows. In section~\ref{sec:DD}
we set the notation for DD and the annual modulation.  In section~\ref{sec:DAMA} we review the case of DAMA modulation in the context of the SHM and we perform a fit to the signal. This section can be skipped by the readers familiar with the DAMA signal. Section~\ref{sec:bound} encodes the main results of this paper, where we derive
lower bounds on the product of the local DM density and the
scattering cross section that are independent of the DM
velocity distribution and are valid for annually modulated signals. We do this in two ways: in section~\ref{sec:bounda} by making use of the fact that constant rates need to be larger than modulations, and in section~\ref{sec:boundb} by using an expansion of the rate on the Earth's velocity \cite{HerreroGarcia:2011aa, HerreroGarcia:2012fu}. As an illustration we apply our halo-independent bounds to the DAMA signal in section~\ref{sec:DAMAbounds}. Finally, we give our conclusions in section~\ref{sec:conclusions}.\footnote{In appendix~\ref{ap:centraleq} we give a detailed derivation of the upper bound on the modulation. In appendix~\ref{ap:bin} we show the binned version of the bounds, while in appendix~\ref{ap:multi} we derive them for multi-target detectors.}

\section{Dark matter direct detection and the annual modulation signal} 
\label{sec:DD}

In this section we review the relevant expressions for DD. We focus on elastic scattering of DM
particles $\chi$ off a nucleus with mass number $A$ depositing a nuclear recoil energy $E_R$.
The differential rate for a 
detector with $n$ different target nuclei is given by:
\begin{equation}\label{eq:R}
  \mathcal{R}(E_R, t) = \frac{\rho_\chi}{m_\chi}\,\sum_{A=1}^n \frac{f_A}{m_A}
  \int_{|\vec{v}| > v_m^A} d^3 v  \, v f_{\rm det}(\vec{v}, t) \frac{d \sigma_A}{d E_R}(v)  \,,\qquad v_m^A=\sqrt{\frac{m_A E_{R}}{2 \mu_{\chi A}^2}}\,,
\end{equation}
where $\rho_\chi$ is the local DM mass density and $v_m^A$ is the minimal
velocity of the DM particle required for a recoil energy $E_{R}$, with $\mu_{\chi A}$ being the reduced mass of the DM--nucleus system. $f_A$ is the mass fraction of the detector nucleus labeled by A. For single-target detectors, there is just one contribution and thus the sum over $A$ is absent.  $f_{\rm det}(\vec{v}, t)$ describes the DM velocity distribution in the detector rest frame, with the normalization 
$\int  d^3 v \, f_{\rm det}(\vec{v}, t) =1$. The velocity
distributions in the rest frames of the detector, the Sun and the
galaxy are related by 
$
f_{\rm det}(\vec{v},t) = f_{\rm Sun}(\vec{v} +
\vec{v}_e(t))=f_{\rm gal}(\vec{v} + \vec{v}_s+\vec{v}_e(t)) \,, 
$
where $\vec{v}_e(t)$ is the velocity vector of the Earth relative to
the Sun and $\vec{v}_s$ is the velocity of the Sun relative to the
galactic halo. 

We will concentrate on elastic
spin-independent (SI) and spin-dependent (SD) contact interactions, where the differential scattering
cross section $d\sigma_A(v)/dE_R$ scales as $1/v^2$, and we will take equal couplings of the DM to neutrons and protons. Then the SI cross
section becomes\footnote{In our notation ``A" specifies the nucleus and represents its mass number.} 
\begin{equation}\label{eq:CS}
  \frac{d\sigma_A}{dE_R}(v) = \frac{m_A \sigma_{\rm SI} A^2}{2\mu^2_{\chi p} v^2} F^2_A(E_R) \,,
\end{equation}
where $\sigma_{\rm SI}$ is the total DM--proton scattering cross
section at zero momentum transfer, $\mu_{\chi p}$ is the DM--proton
reduced mass, and $F_{A}(E_R)$ is a nuclear form factor.  For SD
interactions there is no $A^2$ enhancement, the form factor is different, and $\sigma_{\rm SD}$ will denote the zero-momentum DM--proton scattering cross section. 

Using eqs.~\ref{eq:R} and~\ref{eq:CS} the event rate can be written in a very compact way as
\begin{align}  
\mathcal{R}(E_R,t) = \sum_{A=1}^n\,\mathcal{R}_A(E_R,t) =\mathcal{C}\,\sum_{A=1}^n\,f_A\,A^2  F_{A}^2(E_R) \,\eta (v_m^A,t ) \,,
\label{eq:R0}
\end{align}
where we defined the detector-independent quantity 
\begin{align} 
\mathcal{C} \equiv  \frac{\rho_\chi \sigma_{\rm SI} }{2 m_\chi\mu_{\chi p}^2} \,.
\label{eq:C}
\end{align}
Due to the time-dependence of the Earth's velocity and its orientation with respect to the Sun's velocity DD signals exhibit annual modulations \cite{Drukier:1986tm, Freese:1987wu}. The time dependence in the
event rate is introduced through
\be\label{eq:deta} 
\eta(v_m^A,t)=\overline \eta (v_m^A) + \delta\eta(v_m^A, t) \,,
\ee
where we have defined the constant contribution as
\be\label{eq:eta} 
\overline \eta(v_m^A) \equiv \int_{v>v_m^A}  d^3 v\, \frac{f_{\rm det} (\vec{v})}{v}\,,
\ee
and the modulated one comes from an expansion to first order in $v_{\rm e}=29.8$ km/s $\ll v_{\rm
sun}\simeq$ 230 km/s:
\be
\delta \eta(v_m^A, t) = A_\eta(v_m^A) \cos 2\pi[t - t_*(E_{R})] \,.
\ee
Here $A_\eta(v_m^A) \equiv 0.5\,[\delta\eta(v_m^A, t_*(E_{R})) - \delta\eta(v_m^A, t_*(E_{R})+0.5)]$, where the phase $t_*(E_{R})$ is the time of year at which the event rate is maximum, and thus $A_\eta(v_m^A)>0$. Notice that observing a sign flip in the modulation amplitude would be a strong evidence that the signal is due to DM. However, the phase and the energy at which the flip may occur depends on the velocity distribution. Moreover, there could also be several sign flips at different energies. For the SHM, $t_*^{\rm SHM}=0.42\,\rm y$, corresponding to June 2nd, and there is a sign flip at $v_{m}^A \approx 210\,\rm km/s$, see for instance ref.~\cite{Freese:2012xd}.

Similarly, the amplitude of the modulated rate is:
\be \label{eq:mod}
\mathcal{M}(E_R) = \frac{1}{2}\,\left[\mathcal{R}(E_R,t_*) -\mathcal{R}(E_R,t_*+0.5)\right]\,,
\ee
which is always positive. 

For a specific detector the number of DM induced events in an energy range $[E_1,E_2]$ is given by 
\begin{equation} \label{Nevents} 
N_{[E_1,E_2]}(t) = M \,T\, \mathcal{C}\, [[ \eta  (v_m^A,t) ]]_{E_1}^{E_2} \,,
\end{equation} 
where $M$ and $T$ are the detector mass and exposure time,
respectively, and we define
\begin{equation}\label{eq:shorthand}
[[ X ]]_{E_1}^{E_2}
\equiv 
\sum_{A=1}^n f_A \,A^2
\int_0^\infty d E_R \, F_{A}^2(E_R)\, G_{[E_1, E_2]}^A(E_R) \, X\,,
\end{equation}
which is an integration over energy of the weighted sum of the target nuclei for the quantity $X(v_m^A)$. Here $G_{[E_1,E_2]}^A(E_R)$ is the detector response function
describing the probability that a DM event with true recoil energy
$E_R$ is reconstructed in the energy interval $[E_1,E_2]$, including efficiencies, 
energy resolution, and quenching factors.

\section{The DAMA annual modulation signal within the SHM} \label{sec:DAMA}


In section~\ref{sec:bound} we will derive halo-independent bounds valid for annual modulation signals, and in order to illustrate their use, we will apply them to the modulation observed by DAMA~\cite{Bernabei:2013xsa}. In the following we review and analyze the interpretation of this modulation within the SHM, and therefore this section can be skipped by those readers familiar with this signal.

The DAMA experiment reports an annual modulation of the rate in their NaI scintillator detector, with a period of one year and a maximum around June 2nd. With an exposure of $1.33\,\rm ton\,y$ the modulation is claimed to be compatible with DM at a very high statistical significance ($9.3\sigma$) \cite{Bernabei:2013xsa}. The  amplitude  of the annual modulation observed in the $[2,6]\, \rm keVee$ range is $\mathcal{M}_{[2,6]}=(0.0112 \pm 0.0012)\, \rm counts/keVee/kg/day$ \cite{Bernabei:2013xsa}, while above this range the data is consistent with no annual modulation. As it is well known, the signal can be explained under standard assumptions regarding the DM halo (SHM) and WIMP interactions.  However, the DM interpretation of the modulation is strongly disfavored by other experiments independently of the halo, both for elastic SI and SD interactions \cite{HerreroGarcia:2012fu}, and for inelastic interactions \cite{Bozorgnia:2013hsa}. Despite this, we will use the DAMA signal in the following to illustrate the use of the lower bounds derived in section~\ref{sec:bound}. Notice that if DAMA were not due to DM, current limits (for instance by LUX~\cite{Akerib:2013tjd}) imply that observing annual modulations will be very hard in the future (see for instance ref.~\cite{Bozorgnia:2014dqa}).

First let us discuss which is the dark mass range consistent with the data for the SHM. There is no observation of a sign flip and all the observed modulations in the $[2,6]\, \rm keVee$ range are positive, which means that for the SHM they are above the sign flip, implying that $v_m\gtrsim210\,\rm km/s$. Therefore, in order for the modulation of the first bin to be positive, if the scattering is on sodium, from eq.~\eqref{eq:R} the DM mass needs to be smaller than $\sim 30\,\rm GeV$, while if the scattering is on iodine, it needs to be smaller than $\sim90\,\rm GeV$. In addition, there are also lower bounds on the DM masses in order for the DM particles to be energetic enough to deposit recoil energies above threshold. This means that if the scattering is on iodine, DM masses below $\sim 20\,\rm GeV$ cannot produce observable recoils. On the contrary, for $m_\chi\gtrsim 20$ GeV and SI interactions, iodine typically dominates the rates due to the $A^2$ enhancement.\footnote{For this reason lowering the threshold of the DAMA experiment would in principle allow to distinguish between the two SI solutions, the $\sim10$ GeV DM mass, for which a lower threshold would imply an enhanced rate due to scatterings on iodine, and the $\sim80$ GeV one, which would show a phase flip in the lowest energy bin, see ref.~\cite{Kelso:2013gda}.} Furthermore, by demanding that there are observable recoils in the $6$ keVee range, we obtain that $m_\chi\gtrsim 8\, (30)$ GeV for scattering on Na (I). 

\begin{figure}
	\centering
	\includegraphics[width=0.65\textwidth]{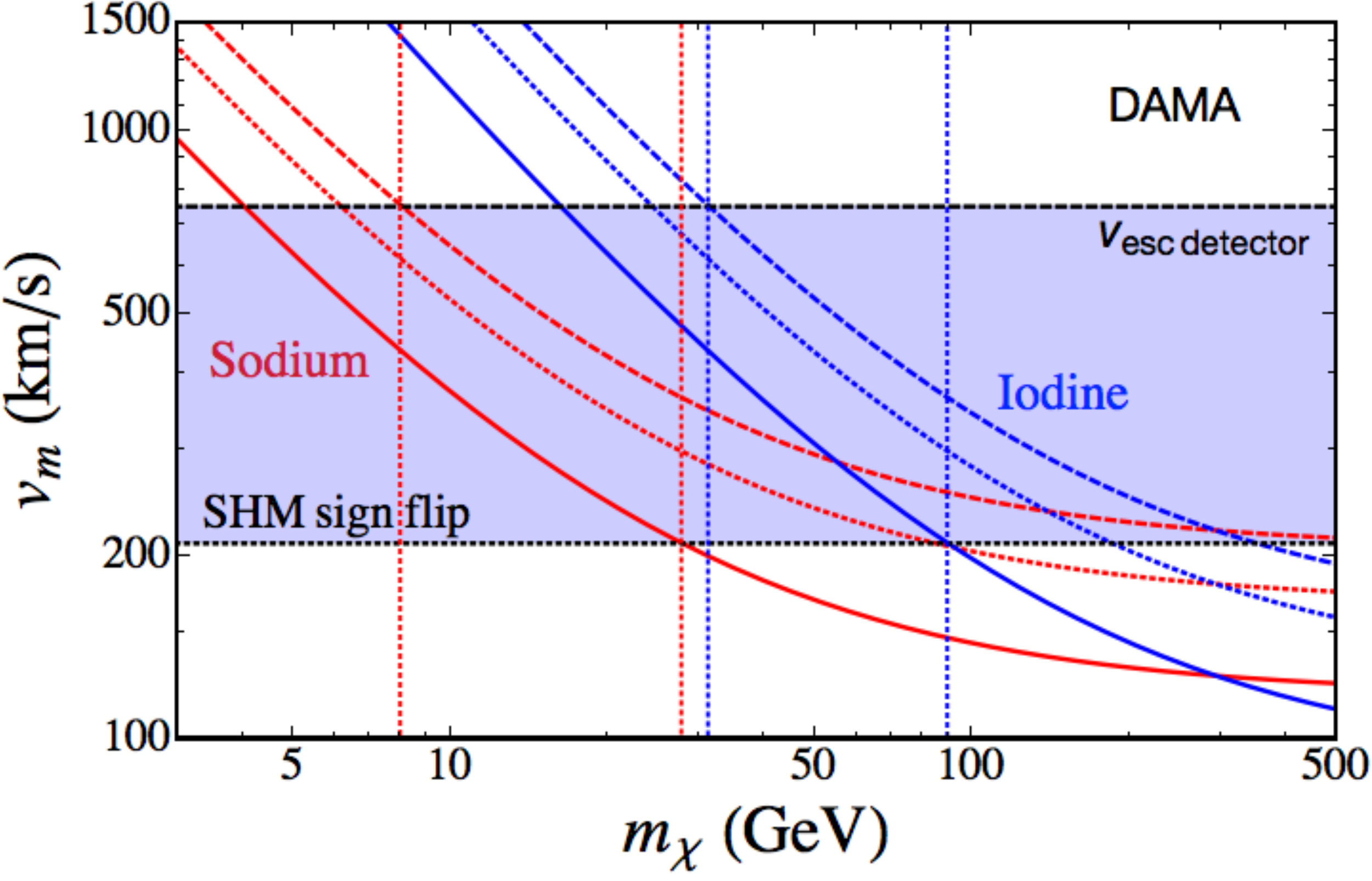}
	\caption{Minimum velocity $v_m(E_R)$, eq.~\eqref{eq:R}, versus the DM mass shown for sodium (red) and iodine (blue) for recoil energies relevant for the DAMA experiment: $E_R=2,\,4\,,6\,\rm  keVee$ (from bottom to top as solid, dotted and dashed curves). It is shown as dotted black the $v_m$ below which the modulation is negative for the SHM, and as dashed black the typical escape velocity in the detector rest frame. The shaded area encoded between the dotted vertical red (blue) lines is the allowed mass range within the SHM for scattering on sodium (iodine).} \label{DAMAmass}
\end{figure}

This is illustrated in figure~\ref{DAMAmass}, where we plot the minimum velocity $v_m(E_R)$, eq.~\eqref{eq:R}, versus the DM mass for sodium (red) and iodine (blue) for recoil energies relevant for the DAMA experiment: $E_R=2,\,4\,,6\,\rm  keVee$ (from bottom to top as solid, dotted and dashed curves). We also show as dotted black the minimum velocity below which the modulation flips sign for the SHM, and as a dashed black line the typical escape velocity in the detector rest frame. The shaded area encoded between the dotted vertical red (blue) lines shows the allowed mass range $8 \lesssim m_\chi \,(\rm GeV)\lesssim 30$ ($30 \lesssim m_\chi \,(\rm GeV) \lesssim 90$) for scattering on sodium (iodine) within the SHM. For more complicated haloes with streams or debris flows the phase of the modulation will be different (see for instance ref.~\cite{Freese:2012xd}), and so will the allowed mass range. 

In order to analyze more quantitatively DAMA's modulation within the SHM, we perform a $\chi^2$ fit to the data in the [2,6] keVee energy range, both for SD and SI elastic interactions. For our numerical analysis we take equal couplings to protons and neutrons, both for SI and SD,\footnote{Regarding SD, both $\rm ^{23}Na$ and $\rm ^{127}I$ are dominated by the spin of the protons. Therefore, suppressed couplings to neutrons help to reduce the inconsistency with other null-result experiments that have nuclei dominated by the spin of the neutrons, such as $\rm ^{73} Ge,\, ^{129} Xe\,\,\rm and\, ^{131}Xe$, see for instance ref.~\cite{Kopp:2009qt}.} and quenching factors $q_{\rm Na} =
0.3$ for sodium and $q_{\rm I} =0.09$ for iodine. For the SI form factor we use the Helm
parameterization, $F(E_R) = 3 e^{-q^2
s^2/2} [\sin(q r)-q r\cos(q r)] / (q r)^3$, with $q^2 = 2 m_A E_{R}$ and $s
= 1$~fm, $r = \sqrt{R^2 - 5 s^2}$, $R = 1.2 A^{1/3}$~fm. For SD we take the structure functions from ref.~\cite{Bednyakov:2006ux}.  For the SHM we use a Maxwellian velocity distribution with mean velocity $\bar v = 220$~km/s truncated at the galactic escape velocity $v_\mathrm{esc} = 550$~km/s, and a local energy density $\rho_\chi=0.4\,\rm GeV\,cm^{-3}$, see ref.~\cite{Read:2014qva} for a recent review.

We use the latest results from figure 8 of ref.~\cite{Bernabei:2013xsa}. As expected from the literature, there are two minima: for the large DM mass solution the scattering is predominantly on iodine, while for the low DM mass one it is on sodium. The best-fit values are shown in table~\ref{tab:DAMAfit}. In all cases we obtain a good fit, with $\chi_{\rm min}^2/{\rm dof}<1.5$, where dof stands for the number of degrees of freedom, which is $6$ ($8$ data points minus $2$ parameters). Notice that for SI interactions, for the large DM mass solution the scattering is predominantly on iodine and therefore the cross section is much lower than in the case of low DM masses (scattering on sodium), due to the larger $A^2$ enhancement of the former. For SD both solutions imply roughly the same cross sections.\footnote{This behaviour can be clearly seen in figure~\ref{DAMAmulti} of appendix~\ref{ap:multi}, where we show in the same plot both solutions in blue (dotted for scattering on Na, dashed on I, and solid on both), for SI (top) and SD (bottom), together with the confidence level regions.} Our results agree with those present in the literature, see for instance refs.~\cite{Kopp:2009qt, Schwetz:2011xm, DelNobile:2015lxa}.

\begin{table}
\centering
\begin{tabular}{|c|c|c|c|c|c|c|}
\hline 
 & \multicolumn{1}{c}{} & \multicolumn{1}{c}{\bf I} &  & \multicolumn{1}{c}{} & \multicolumn{1}{c}{\bf Na} & \tabularnewline
\hline 
\hline 
& $m_\chi\,(\rm GeV)$& $\sigma_{\rm SI/SD}\, (\rm cm^2)$& $\chi_{\rm min}^2/{\rm dof}$ & $m_\chi\,(\rm GeV)$& $\sigma_{\rm SI/SD}\, (\rm cm^2)$& $\chi_{\rm min}^2/{\rm dof}$\tabularnewline
\hline 
\bf SI & $79.4$ & $1.1\cdot10^{-41}$ & $7.7/6$ & $12.6$& $1.8\cdot10^{-40}$&$8.3/6$\tabularnewline
\hline 
\bf SD & $63.1$ & $5.0\cdot10^{-37}$ & $7.9/6$ & $12.6$& $6.3\cdot10^{-37}$&$8.7/6$\tabularnewline
\hline 
\end{tabular}
\caption{\label{tab:DAMAfit} Results of the DAMA fit to the SHM, done for SI and SD interactions, assuming equal couplings to protons and neutrons. We use $v_\mathrm{esc} = 550$~km/s, $\rho_\chi=0.4\,\rm GeV\,cm^{-3}$ and the quenching factors $q_{\rm Na} =0.3$ for sodium and $q_{\rm I} =0.09$ for iodine.
}
\end{table}

\begin{figure} [h]
	\centering
	\includegraphics[width=0.48\textwidth]{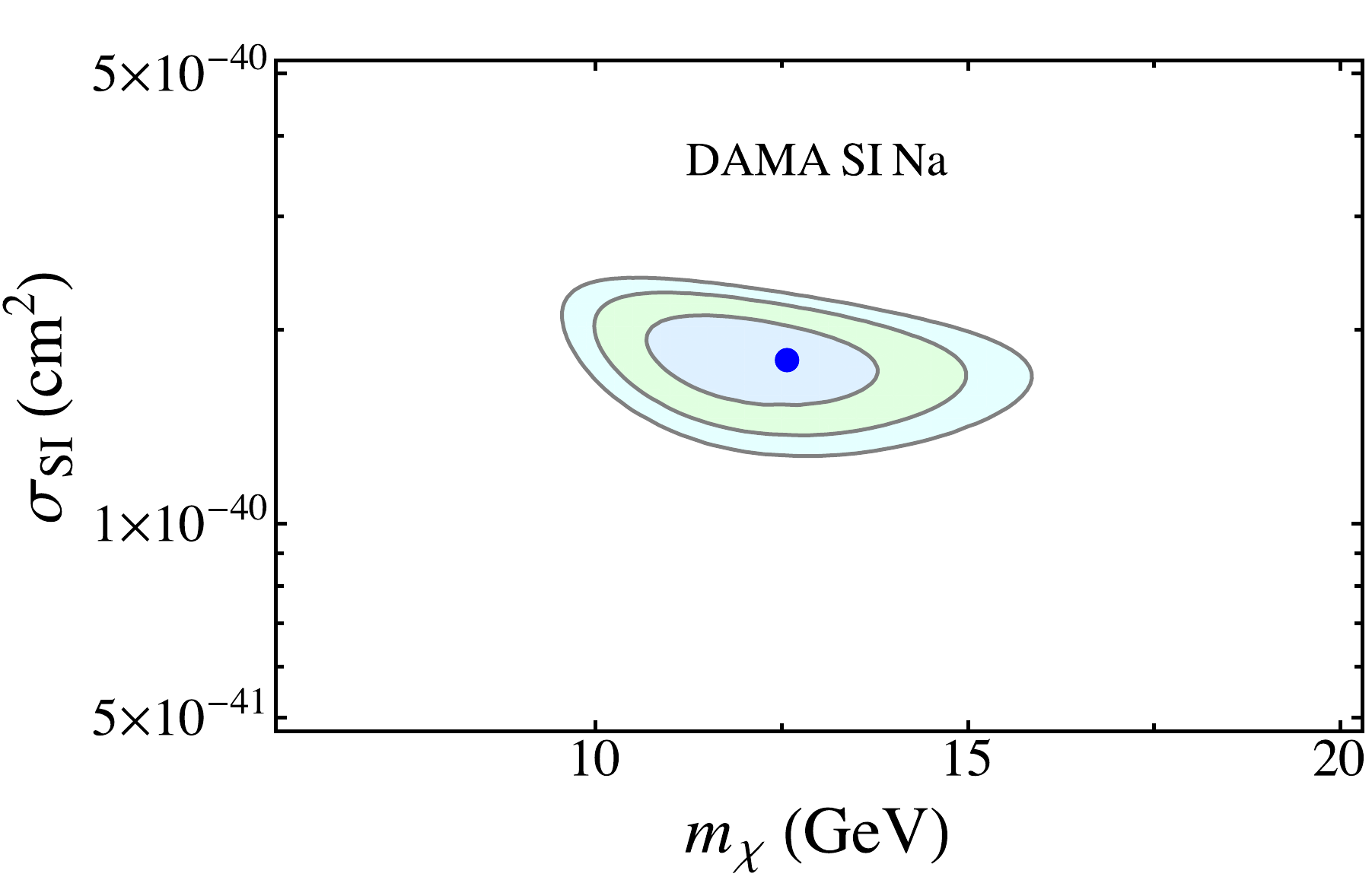}~~\includegraphics[width=0.48\textwidth]{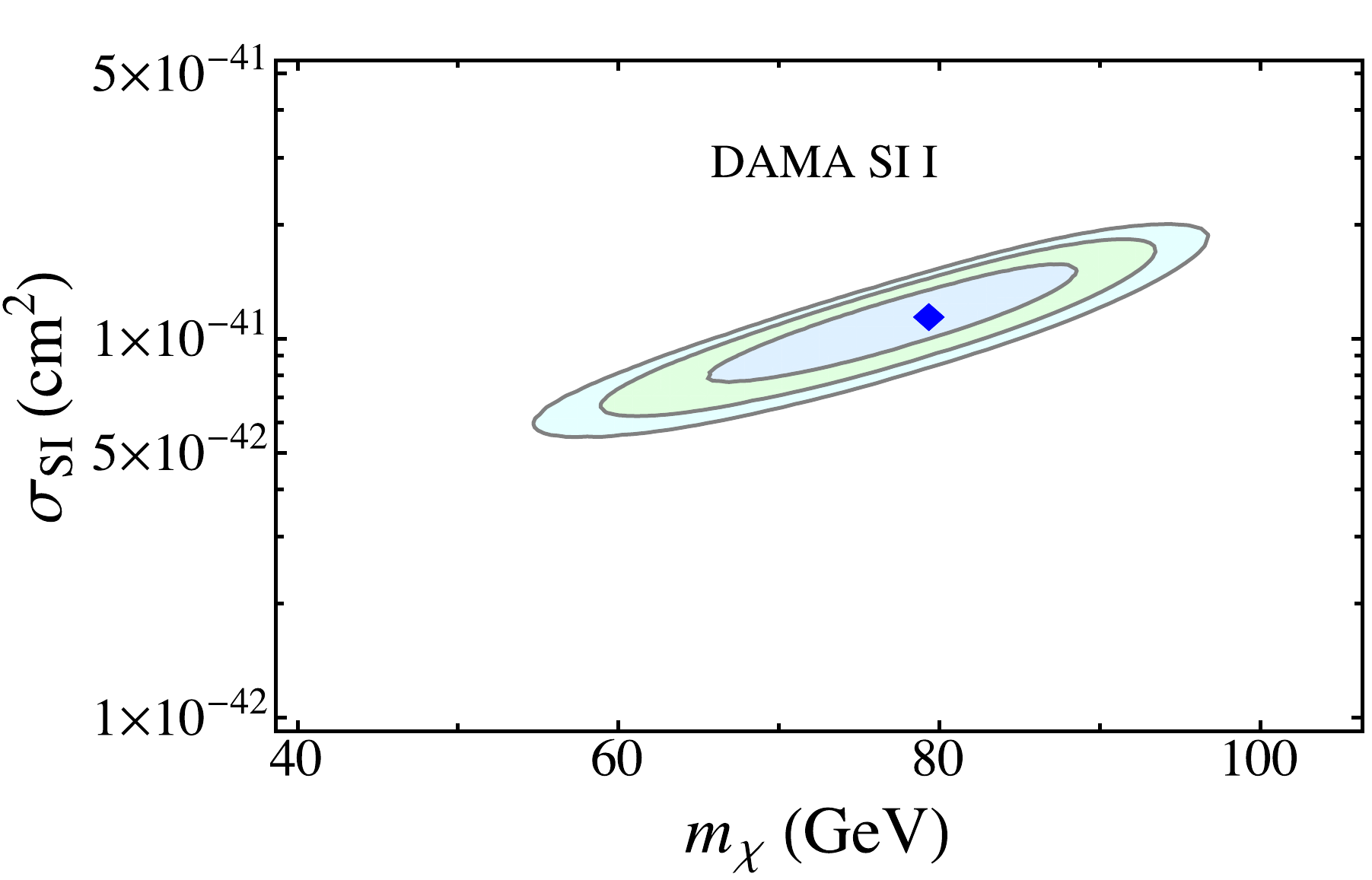}\\
	\includegraphics[width=0.48\textwidth]{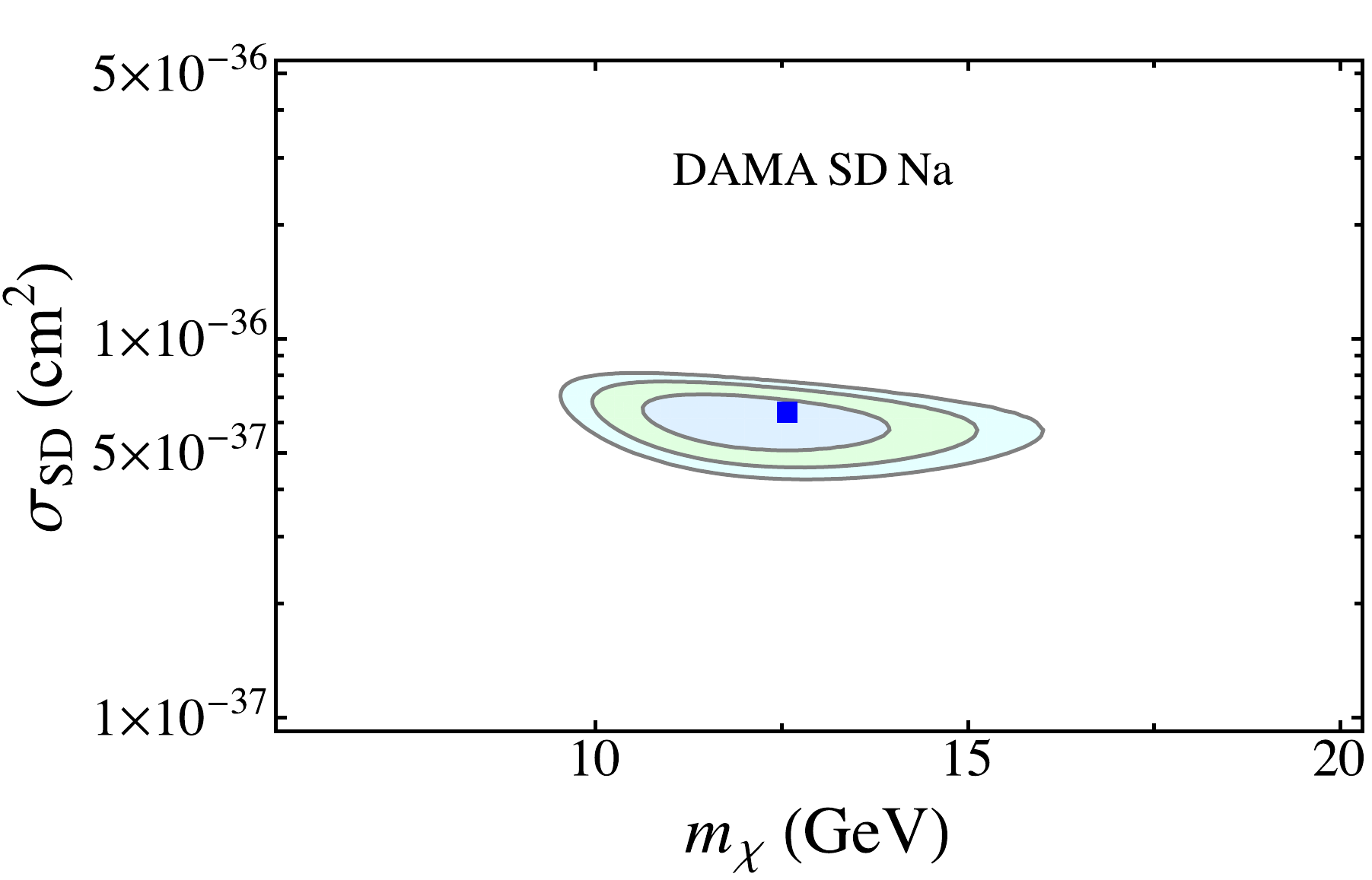}~~\includegraphics[width=0.48\textwidth]{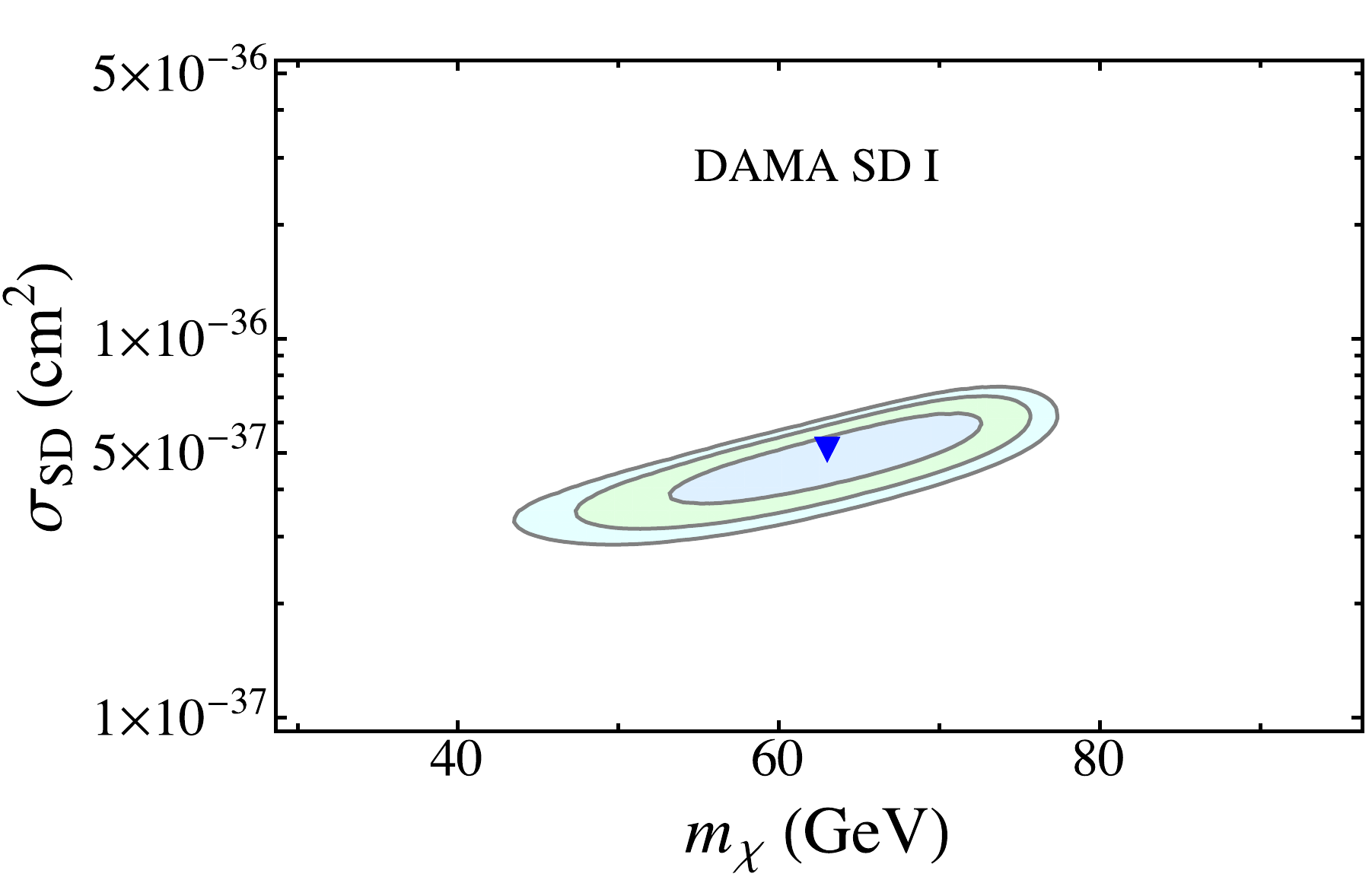}
	\caption{SHM best-fit regions for the DAMA data for SI (top) and SD (bottom), and scattering on Na (left) and I (right). The $\Delta\chi^2=2.3,5.99,9.21$ contours corresponding in two dimensions to a CL of $68.27, 95, 99 \%$ are shown in blue, green and light blue, respectively, together with the best-fit points, depicted with blue marks.} \label{DAMAbf}
\end{figure}

In figure~\ref{DAMAbf}, we show the allowed $90\%\,\rm CL$ parameter space for the SHM as solid blue lines, together with the $\Delta\chi^2=2.3,5.99,9.21$ contours (in blue, green, light blue respectively) corresponding in two dimensions to a CL of $68.27, 95, 99 \%$ respectively. This is done for SI (top) and SD (bottom), and we show the low (high) mass solution in the left (right), which correspond to scattering on Na (predominantly on I). We also plot the best-fit points with blue marks.

\section{Halo-independent lower bounds on 
$\rho_\chi \sigma_{\rm SI/SD}$ for annual modulations}  \label{sec:bound}

In this section we derive halo-independent bounds valid for annual modulation signals. It encodes the most important results of this work. 
\subsection{Using that the constant rate is larger than the annual modulation} \label{sec:bounda}
An upper bound on the halo integral $\overline\eta(v_m)$\footnote{To alleviate notation, in the following we will drop the $A$ superscipt for the minimum velocity, $v_m^A$, see eq.~\eqref{eq:R}. We will show it explicitly when it is crucial to illustrate its dependence on the nucleus.} can be easily obtained using the normalization condition of the velocity distribution as well as the definition of $\overline\eta(v_m)$, eq.~\eqref{eq:eta}, as was done in refs.~\cite{Kavanagh:2012nr, Feldstein:2014ufa, Blennow:2015gta}:
\begin{align}  
1 &\equiv \int_0^\infty d^3v f_{\rm det}(\vec{v}) \equiv \int_0^\infty \overline\eta(v) dv \\
&\geq\overline  \eta(v_1)\, v_1+ \int_{v_1}^{v_2} dv \, \overline\eta(v)\, \label{eq:bound2} \\
&\geq\overline \eta(v_1)\, v_1  \,, \qquad  \label{eq:bound1}
\end{align}
where from the first to the second line we used that below the threshold $v_1$ the minimum is obtained for constant $\overline{\eta}(v)=\overline{\eta}(v_1)$, as it is a monotonously decreasing function, and in the last line we dropped the second term.

From these expressions an upper bound on the constant rate can be derived~\cite{Blennow:2015gta}. In order to have a bound valid for annual modulation signals, one can simply use that the constant rate is larger than the amplitude of the modulation:
\be \label{eq:Asmallereta} 
\overline\eta(v_m)\geq A_\eta(v_m)\,.
\ee
If the spectrum is not measured, one can derive a bound in terms of the measured number of modulated events in an energy interval $[E_1,E_2]$ by combining eq.~\eqref{Nevents} with eqs.~\eqref{eq:bound1} and~\eqref{eq:Asmallereta}. Using the definition of $\mathcal{C}$, eq.~\eqref{eq:C}, one gets:
\be \label{eq:bound1mod} 
\rho_\chi \sigma_{\rm SI} \geq 
\frac{2 \,m_\chi\,\mu_{p}^2}{MT\,[[ 1/v_m^A]]_{E_1}^{E_2} } 
\, N_{[E_1, E_2]}^{\rm mod} \, \,\qquad\qquad \rm ``Events \,bound",
\ee
where $[[1/v_m^A]]_{E_1}^{E_2}$ is defined in
eq.~\eqref{eq:shorthand}. Notice that this bound can be used for multi-target experiments.

If the spectrum is measured, one can obtain a stronger bound using eq.~\eqref{eq:bound2} instead of eq.~\eqref{eq:bound1}. In this case, for a single-target experiment with perfect energy resolution, for fixed DM mass, a measurement of the spectrum $\mathcal{M}(E_R)$ allows a determination of the halo integral via eqs.~\eqref{eq:R0} and~\eqref{eq:mod}:
\begin{align}\label{eq:eta_obs}
A_\eta (v_m) =
\frac{\mathcal{M}(E_R)}{\mathcal{C}\,A^2  F_{A}^2(E_R)} \,.
\end{align}
Then, in combination with eq.~\eqref{eq:bound2}, the following bound can be obtained: 
\begin{align}\label{eq:bound2mod}
  \rho_\chi \sigma_{\rm SI} \ge \frac{2 m_\chi \mu^2_{\chi p}}{A^2} 
  \left( v_1 \frac{\mathcal{M}(E_1)}{F_{A}^2(E_1)} + 
         \int_{v_1}^{v_2}dv \frac{\mathcal{M}(E_R)}{F_{A}^2(E_R)}  \right) \,\,\qquad\qquad \rm ``Spectrum\,bound".
\end{align}
If a DD experiment observes an annual modulation spectrum, eq.~\eqref{eq:bound2mod} provides a lower bound on the product $\rho_\chi \sigma_{\rm SI}$ which is independent of the local DM
velocity distribution. In ref.~\cite{Blennow:2015gta} bounds valid for constant rates analogous to those of eqs.~\eqref{eq:bound1mod} and~\eqref{eq:bound2mod} were derived. Notice that if a sign flip is observed, a better approach is to apply the bound for the region below and above the sign flip separately, in order to obtain stronger bounds.

If the detector has different elements, in order to use eq.~\eqref{eq:bound2mod} one needs to assume that one of them gives the dominant contribution. The bound will be stronger in general for the lowest mass nuclei in the case of SI, due to the $A^2$ suppression, but so will the SHM prediction, and therefore we expect the ratio of the bound and the SHM to remain approximately the same for different nuclei. One also needs to take into account that, although for SI interactions the heaviest element typically dominates the rate due to the $A^2$ enhancement, for low DM masses the scattering on the heavy nuclei may not be possible, as there are no DM particles with speeds larger than the escape velocity in the detector rest frame, $\sim 750 \,\rm km\,s^{-1}$. We will illustrate these features for the DAMA data in section~\ref{sec:DAMAbounds}. In appendix~\ref{ap:multi} we generalize eq.~\eqref{eq:bound2mod} to the case of multi-target detectors. 

\subsection{Using upper bounds on the annual modulation} \label{sec:boundb}
One a can derive stronger tests for the case that the spectrum is measured by using the results of refs.~\cite{HerreroGarcia:2011aa, HerreroGarcia:2012fu}, where upper bounds on the annual modulation signal in terms of the constant rate were obtained by doing a first order expansion of the rate on the Earth's velocity.

Lets consider a spectral measurement of both the modulation $\mathcal{M}(E_R)$ and the constant rate $\mathcal{R}(E_R)$ in the energy
range $[E_1,E_2]$, which for a given DM mass can be related to a velocity interval $[v_1, v_2]$ via eq.~\eqref{eq:R}. By doing an expansion on the Earth's velocity and some mild assumptions about the halo (see below) one can derive the following upper bounds on the modulation amplitude~\cite{HerreroGarcia:2011aa, HerreroGarcia:2012fu}:
\begin{align}
\int_{v_{1}}^{v_{2}} \negthickspace \negthickspace dv\, A_\eta(v) &
\leq  v_e\left(\overline \eta(v_{1}) +
\int_{v_{1}}^{v_{2}} \negthickspace \negthickspace
dv \frac{\overline\eta(v)}{v} \right) \qquad &\rm ``General \,halo", 
\label{eq:bound_gen2}\\
\int_{v_{1}}^{v_{2}} \negthickspace \negthickspace dv\, A_\eta(v)  &
\leq  \sin\alpha \, v_e \, \overline \eta(v_{1}) \qquad &\rm ``Symmetric\, halo".
\label{eq:bound_spec2}
\end{align}
For the ``General halo" it is assumed that the velocity distribution is constant in time on the scale of 1 year (i.e., that all the time dependence of the rate comes from the velocity of the Earth) and that it is constant in space on the scale of the size of the Sun-Earth distance. Notice that halo substructures like streams with velocities larger than $v_e$ are covered by the bounds. The ``Symmetric halo" further assumes that there is a preferred direction for the DM flow, specified by $\alpha$, the angle between the DM flow and the orthogonal to the ecliptic. The most conservative bound is obtained for $\sin\alpha =
1$, which corresponds to a DM stream parallel to the ecliptic. However, it may happen that the DM
velocity is aligned with the motion of the Sun, and therefore $\sin\alpha \simeq 0.5$. This is the case of isotropic velocity distributions and, up to a small correction due to the peculiar velocity of the Sun, it also holds for triaxial halos or a dark-disc. For the ``General halo" the bound is independent of the phase, which is free. However, for the ``Symmetric halo" the phase is independent of $v_m$ (and therefore independent of $E_R$), and fixed. The SHM is included in this last case, with $\sin\alpha \simeq 0.5$ and the phase being equal to June 2nd. As for the case of the ``Spectrum bound", eq.~\eqref{eq:bound2mod}, if a sign flip in the modulation is observed in the velocity range considered, one may apply the bound for the region below and above the sign flip separately, in order to get stronger bounds. More details regarding the different assumptions of the bounds can be found in the original ref.~\cite{HerreroGarcia:2011aa} (and also in refs.~\cite{HerreroGarcia:2012fu,Bozorgnia:2013hsa}).

From the above discussion it is clear that in order to decide whether to apply the ``General halo" or the ``Symmetric halo" one can use the information encoded in the phase. First of all, one needs to do a fit keeping as free parameters the modulation amplitude, the period and the phase. If data disfavours a constant phase across the different energy bins, this invalidates the assumption of a ``Symmetric halo" and one is forced to use the ``General halo". Regarding the polar angle $\alpha$, unfortunately, the phase of the modulation does not carry information about it, and different values of $\alpha$ will give the same phase. It only depends on the azimuthal angle within the ecliptic, $\phi$, on which however the bounds do not depend. This is easy to visualize if one imagines a stream parallel to the ecliptic, with $\alpha=\pi/2$. In this case it is clear that the phase is completely determined by the angle $\phi$ at which the Earth's velocity and the stream direction are aligned. Notice that the amplitude of the modulation is proportional to $\sin\alpha$, but, unfortunately, its value cannot be disentangled without specifying the halo.

In order to check the consistency of a DD signal one can apply these bounds if both the modulation and the constant rate have been measured~\cite{HerreroGarcia:2011aa}. In some cases one may not be able to separate the background from the DM contribution in the measured constant rate, as is the case of DAMA, but it is clear that if the inequalities are violated for an unknown constant background plus DM rate, they will also be for just the DM component, and the conclusion (the exclusion of the DM as an interpretation of the modulation) will still hold. However, if the bounds are fulfilled for some unknown constant background plus DM rate, they may well be violated for just the DM rate. In these cases, in order to check if the modulation observed in one experiment is consistent, one can compare it with upper limits on the rate from a different experiment \cite{HerreroGarcia:2012fu,Bozorgnia:2013hsa}. Interestingly, as we will show now, one can obtain upper bounds that do not depend in any way on the constant rate nor on its upper limits. In combination with other measurements, these set halo-independent constraints on the DM models~\cite{Blennow:2015gta}. 

By applying eq.~\eqref{eq:bound2} in the ``General halo", eq.~\eqref{eq:bound_gen2}, we can obtain an upper bound on the modulation (see appendix~\ref{ap:centraleq} for the derivation):\begin{equation}
\int_{v_{1}}^{v_{2}} \negthickspace \negthickspace dv A_\eta(v)
\leq  v_e\left(\frac{2}{v_1} -\frac{1}{v_2} \right) \,.
\label{eq:boundboth_gen}
\end{equation}

Similarly, for the ``Symmetric halo", eq.~\eqref{eq:bound_spec2}, we obtain:
\begin{equation}
\int_{v_{1}}^{v_{2}} \negthickspace \negthickspace dv A_\eta(v) \leq  \sin\alpha \, v_e \, \frac{1}{v_1}   \,. 
\label{eq:boundboth_spec}
\end{equation}
Both bounds are written completely in terms of velocities, with no factors containing the constant rate. Therefore, even if just the annual modulations are measured, one can get an upper bound. This is crucial, since, as we argued, some experiments are not able to disentangle constant DM signals from constant backgrounds.

If we now consider a spectral measurement of the modulation such that we can extract $A_\eta$ via eq.~\eqref{eq:eta_obs}, then, using eq.~\eqref{eq:boundboth_gen}, we can get an upper bound on the integrated modulation over the velocity range:
\begin{equation} \label{eq:RAta}
\int_{v_{1}}^{v_{2}} \negthickspace \negthickspace dv \frac{\mathcal{M}(v)}{\,A^2\, F_A^2(E_R)} \leq \mathcal{C}  \,v_e\,\left(\frac{2}{v_1} -\frac{1}{v_2} \right)\,,
\end{equation}
from which we can derive a lower bound in the product of the energy density and the cross-section which does not depend on $f(v)$, and is valid for very generic haloes:
\be \label{eq:final_bound} 
\rho_\chi \sigma_{\rm SI} \geq 
\frac{2 \,m_\chi\,\mu_{p}^2}{A^2}\,\frac{1}{v_e}\left(\frac{2}{v_1} -\frac{1}{v_2} \right)^{-1} \, \int_{v_{1}}^{v_{2}} \negthickspace \negthickspace dv \frac{\mathcal{M}(v)}{F_A^2(E_R)} \, \qquad\qquad \rm ``General\, bound".
\ee
An analogous bound can be obtained for symmetric haloes, eq.~\eqref{eq:boundboth_spec}:\footnote{We keep it in this form of $\Big(\frac{1}{v_1}\Big)^{-1}$ instead of writing $v_1$ for easier comparison with the binned version, see eq.~\eqref{eq:final_boundbbin}  in appendix~\ref{ap:bin}.}
\be \label{eq:final_boundb} 
\rho_\chi \sigma_{\rm SI} \geq 
\frac{2 \,m_\chi\,\mu_{p}^2}{A^2}\,\Big(\frac{1}{v_1}\Big)^{-1}\,\frac{1}{\sin\alpha\, v_e}\, \, \int_{v_{1}}^{v_{2}} \negthickspace \negthickspace dv \frac{\mathcal{M}(v)}{F_A^2(E_R)} \, \qquad \qquad \rm ``Symmetric\, bound".
\ee
If a DD experiment observes an annual modulation, eq.~\eqref{eq:bound1mod} (``Events bound"), eq.~\eqref{eq:bound2mod} (``Spectrum bound"), eq.~\eqref{eq:final_bound} (``General bound") and eq.~\eqref{eq:final_boundb} (``Symmetric bound") provide lower bounds on the
product $\rho_\chi \sigma_{\rm SI}$ which are independent of the local DM velocity distribution. The first two just use that modulations are smaller than constant rates, while the last two are based on an expansion of the rate on the Earth's velocity, and are clearly a factor $\sim v/v_e$ (with $v>v_m$) stronger than the former. These are the main results of this paper. Notice if the detector has different nuclei, in order to use eqs.~\eqref{eq:bound2mod},~\eqref{eq:final_bound} and~\eqref{eq:final_boundb}, one needs to assume that one of them gives the dominant contribution. In appendix~\ref{ap:multi} we generalize these bounds to the case of multi-target detectors.

\subsection{Applying the halo-independent bounds: the case of DAMA} \label{sec:DAMAbounds}

Now we will apply the halo-independent bounds derived in the previous section to the annual modulation observed by DAMA~\cite{Bernabei:2013xsa} (see section~\ref{sec:DAMA} for a review and analysis of DAMA in the context of the SHM). As discussed in section~\ref{sec:boundb}, in order to know if one can apply the ``Symmetric bound" or only the ``General bound" one should perform a fit to the modulation data leaving the amplitude, the period and the phase free in order to determine whether the phase is constant or not across the energy range of the signal. For DAMA these fits have been performed~\cite{Bernabei:2013xsa}, obtaining that the phase is compatible with being constant across different energy bins within $3\sigma$, and equal to June 2nd (see table 4 of ref.~\cite{Bernabei:2013xsa}).

It should be emphasized that the DAMA modulation is compatible with the SHM (and thus with $\sin\alpha \simeq 0.5$), see section~\ref{sec:DAMA}, but it can also be caused for instance, as discussed in section~\ref{sec:boundb}, by a stream with a different polar angle, $\sin\alpha \neq 0.5$, but with the same azimuthal angle $\phi$ as in the SHM (where it is given by the Sun's velocity) such that the phase is June 2nd, and with a stream speed such that the final modulation amplitude has the same value as the one observed by DAMA. Therefore, for DAMA one can use the ``Symmetric halo" for different values of $\alpha$. In this case, the most conservative option is $\sin\alpha = 1$, and if one is interested in analysing a DM flow aligned with the velocity of the Sun one can use $\sin\alpha \simeq 0.5$. In order to illustrate their different strengths we will also show the ``General halo", which does not require a constant phase.\footnote{Notice that the differences in the modulation amplitudes and their errors in the case of fixed (free) phase, c.f. table 3 and 4 of ref.~\cite{Bernabei:2013xsa}, are negligible, and therefore we can use the same data when applying both bounds. However, in general this may not be the case, see for instance the CoGENT analysis of ref.~\cite{HerreroGarcia:2011aa}, which uses the extracted modulation amplitudes for the different assumptions of ref.~\cite{Fox:2011px}.}

In order to apply the bounds to real data we need binned expressions, which are given explicitly in appendix~\ref{ap:bin}. To extract $A_\eta(v_m)$ from the observed modulation amplitude $\mathcal{M}(E_R)$ via eq.~\eqref{eq:eta_obs}, and therefore to apply our bounds, one needs to assume that the scattering occurs  on a particular nucleus. We will assume in the following that the scattering is dominated by either sodium or iodine, which, as we explained before, is a very good approximation, given the large mass hierarchy between them. In appendix~\ref{ap:multi} we generalize the bounds to multi-target experiments and apply them to both nuclei at the same time.

\begin{figure}
	\centering
	\includegraphics[width=0.5\textwidth]{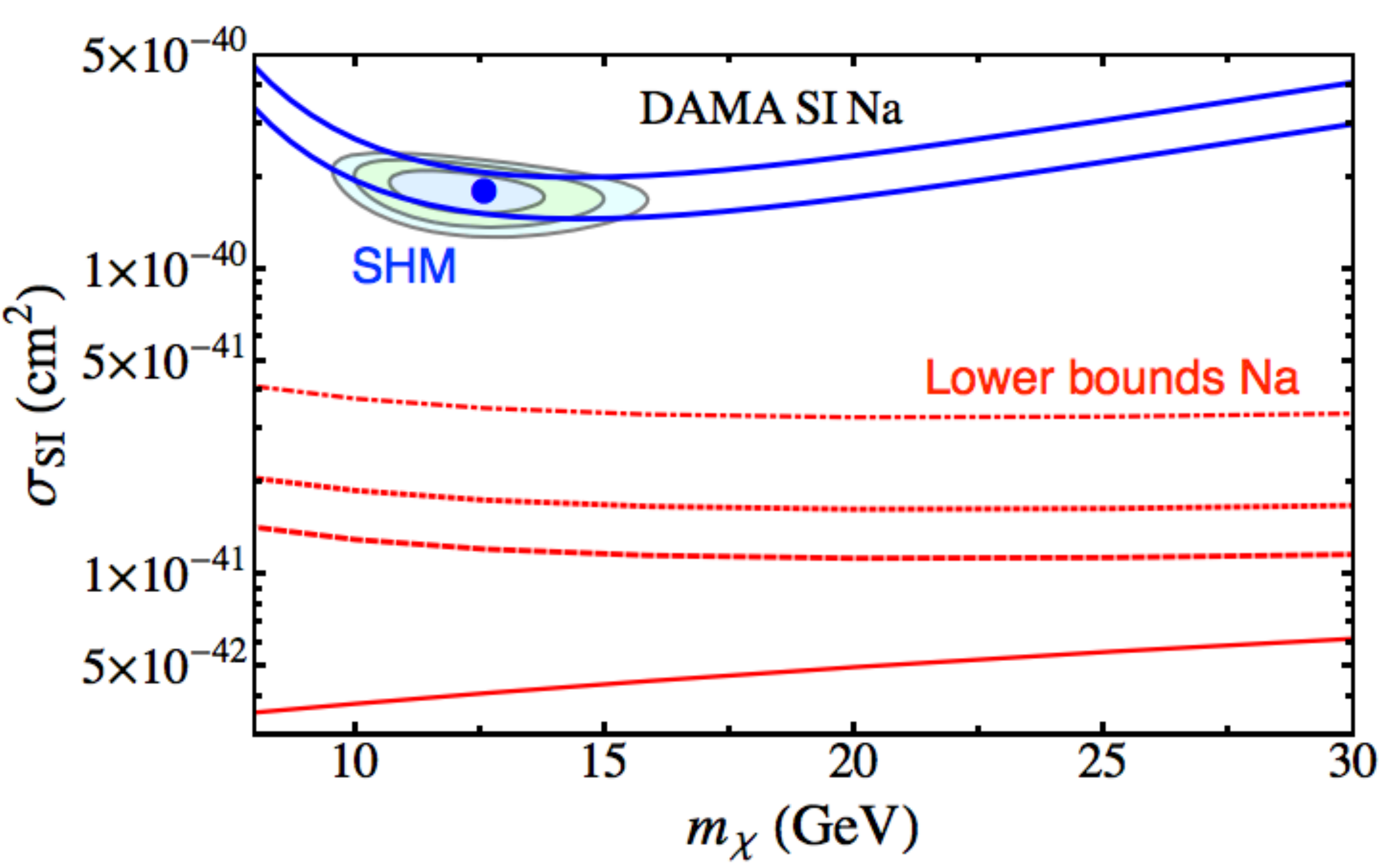}~~\includegraphics[width=0.5\textwidth]{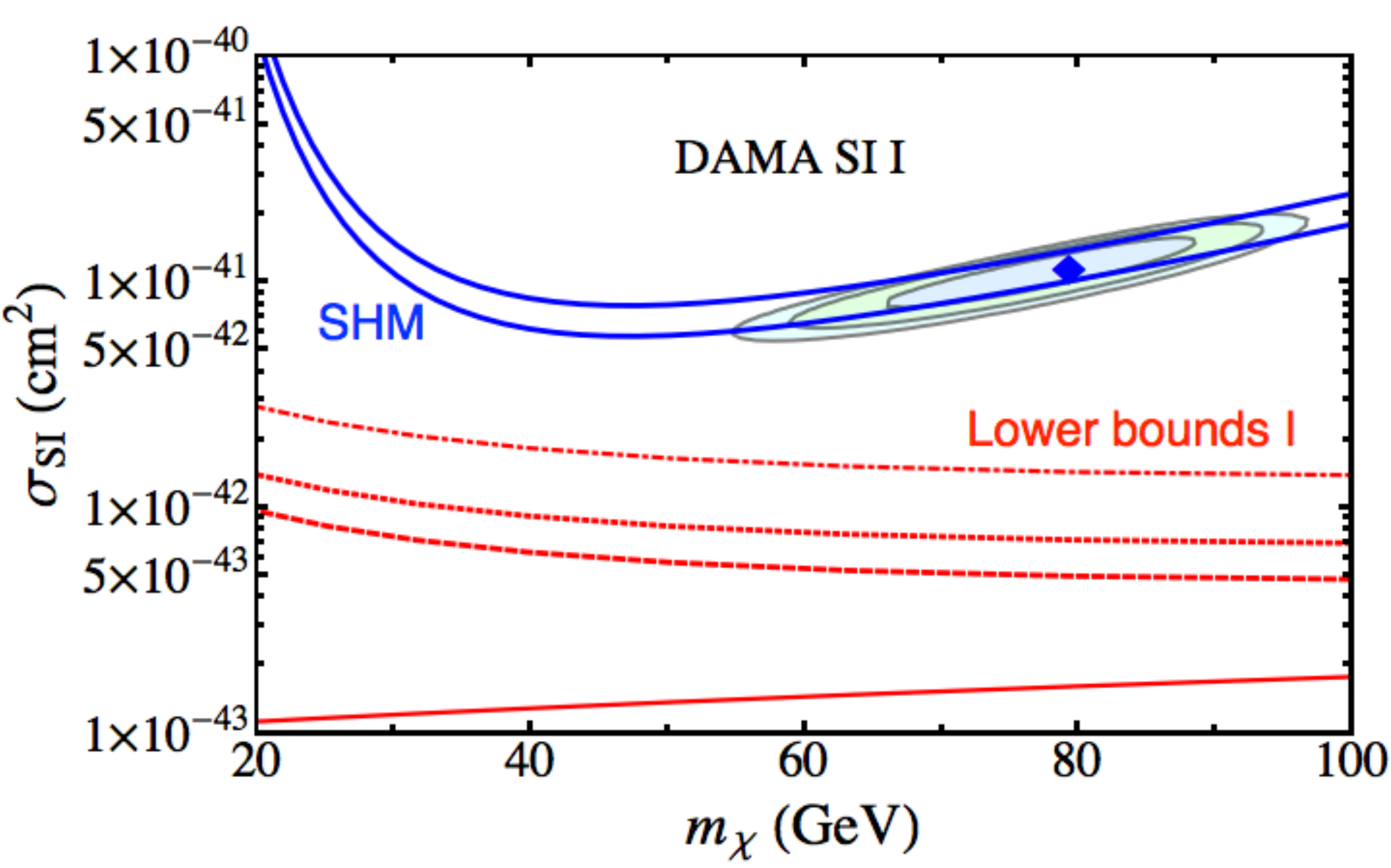}\\
	\includegraphics[width=0.5\textwidth]{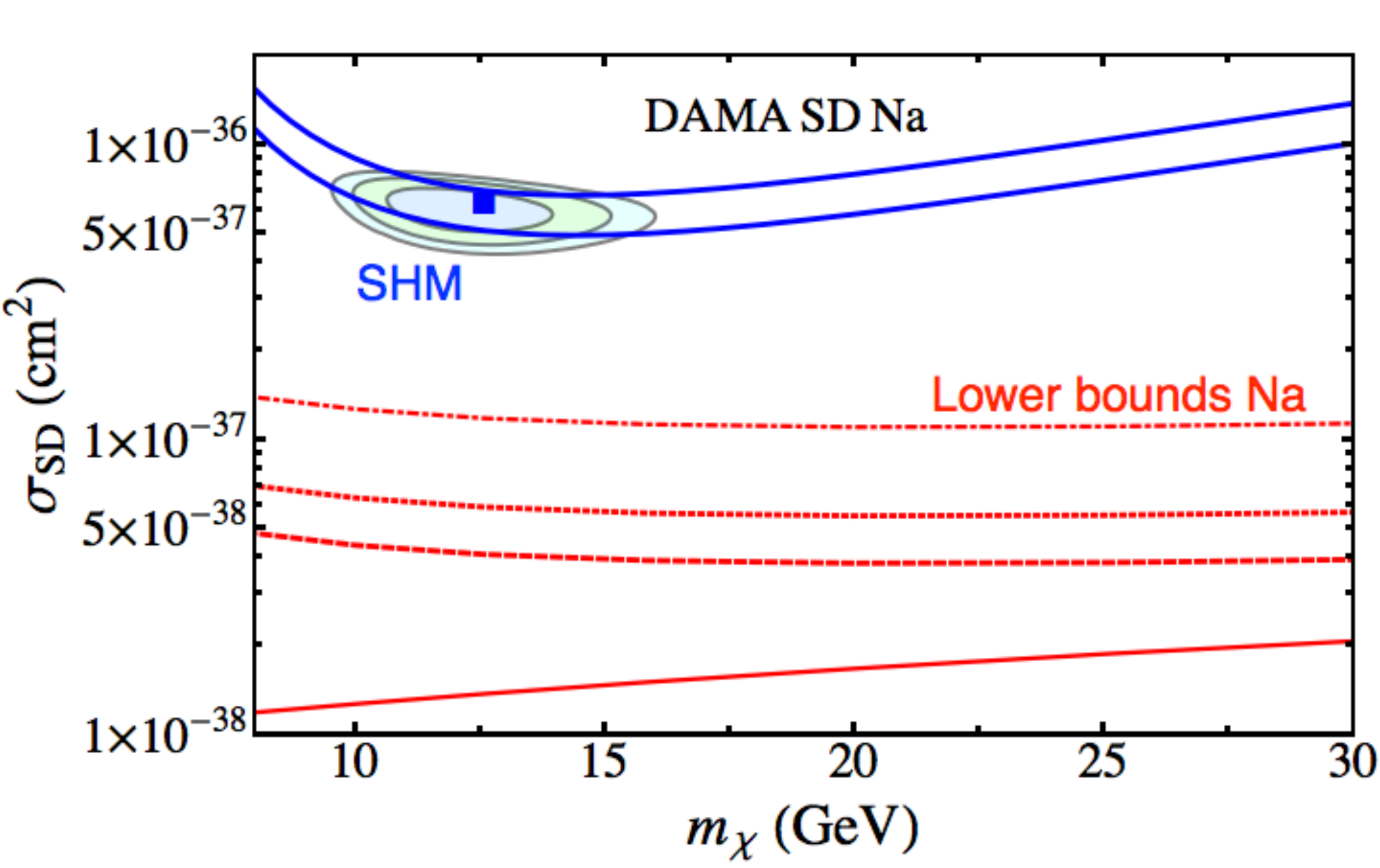}~~\includegraphics[width=0.5\textwidth]{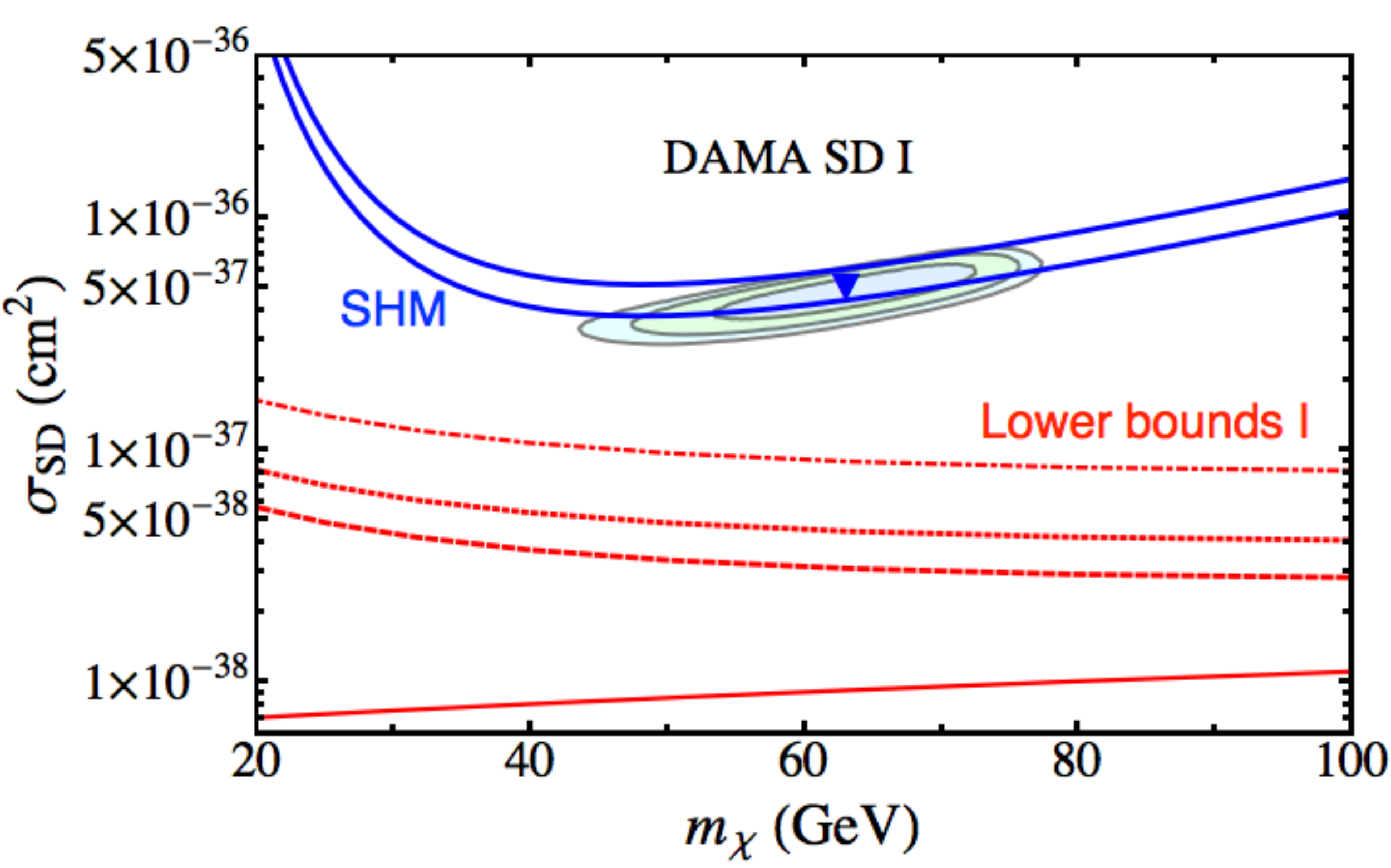}
	\caption{Results for the DAMA data for SI (top) and SD (bottom), and scattering on Na (left) and I (right). The SHM allowed parameter space at the $90\%\,\rm CL$ is encoded between the solid blue lines. The $90\%\,\rm CL$ lower bounds are shown in red from bottom to top: the ``Spectrum bound" of eq.~\eqref{eq:bound2mod} (solid red), the ``General bound" of eq.~\eqref{eq:final_bound} (dashed red) and the ``Symmetric bound" of eq.~\eqref{eq:final_boundb} (dotted red for $\sin \alpha=1$, dotted-dashed red for $\sin \alpha=0.5$. Notice that the latter is a factor of 2 stronger than the former.). The SHM $\Delta\chi^2=2.3,5.99,9.21$ contours corresponding in two dimensions to a CL of $68.27, 95, 99 \%$ are shown in blue, green and light blue, respectively, together with the best-fit points, depicted with blue marks.} \label{DAMA}
\end{figure}

In figure \ref{DAMA} we show the $90\%\,\rm CL$ lower bounds on the scattering cross section, together with the SHM preferred regions, already shown in figure~\ref{DAMAbf}. For definiteness, we use the mass range allowed for the SHM (plotted in blue), see section~\ref{sec:DAMA} for details. From bottom to top, we plot the ``Spectrum bound", eq.~\eqref{eq:bound2mod}, which just uses that $A_\eta\leq\overline{\eta}$, as solid red; we also show the bounds that use the expansion on the Earth's velocity: the ``General bound", eq.~\eqref{eq:final_bound}, as dashed red, and the ``Symmetric bound", eq.~\eqref{eq:final_boundb}, as dotted (dotted-dashed) red for $\sin \alpha=1$ ($\sin \alpha=0.5$, i.e., for the DM flow aligned with the velocity of the Sun). As expected, in all cases the latter bounds are much stronger than the ``Spectrum bound" (solid red), due the $v/v_e$ enhancement (c.f. eqs.~\eqref{eq:final_bound} and~\eqref{eq:final_boundb} to eq.~\eqref{eq:bound2mod}).

One can use these lower bounds on the cross section to compare with LHC limits, with relic abundance constraints or with indirect detection limits~\cite{Blennow:2015gta}. This has to be done within the context of a particular particle physics model and can be used to check its validity.  Another useful and general comparison can be done by combining the lower bound on $\rho_\chi$ for a given DM mass with local density measurements~\cite{Blennow:2015gta}. This combination provides a lower bound on the cross section of any given particle physics model. In the analysis, we use values for $\rho_\chi$ in the range between 0.2 and 0.6~GeV/cm$^3$ (see ref.~\cite{Read:2014qva} for a recent review), and we fix the DM mass to be equal to the best-fit values obtained in the $\chi^2$ fit, see table~\ref{tab:DAMAfit}. In this case one would need to obtain the DM mass by other means, for instance from LHC measurements, from an indirect detection signal (i.e., a gamma ray line) or from more than one DD signal, see for instance refs.~\cite{Drees:2008bv,Kavanagh:2013wba}. Notice that results for different DM masses can be obtained by a simple rescaling of figure~\ref{DAMA}.

\begin{figure}
	\centering
	\includegraphics[width=0.51\textwidth]{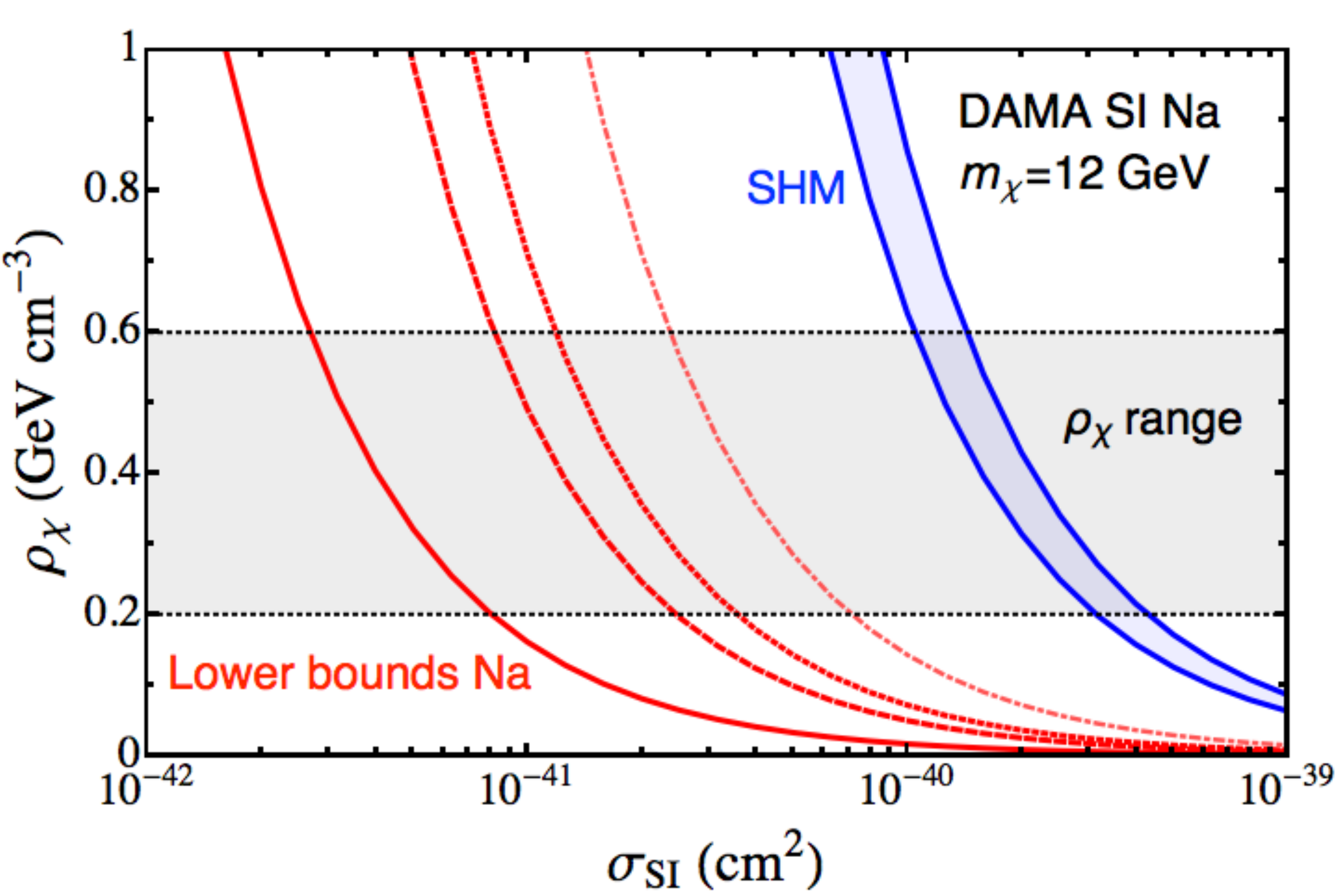}~\includegraphics[width=0.51\textwidth]{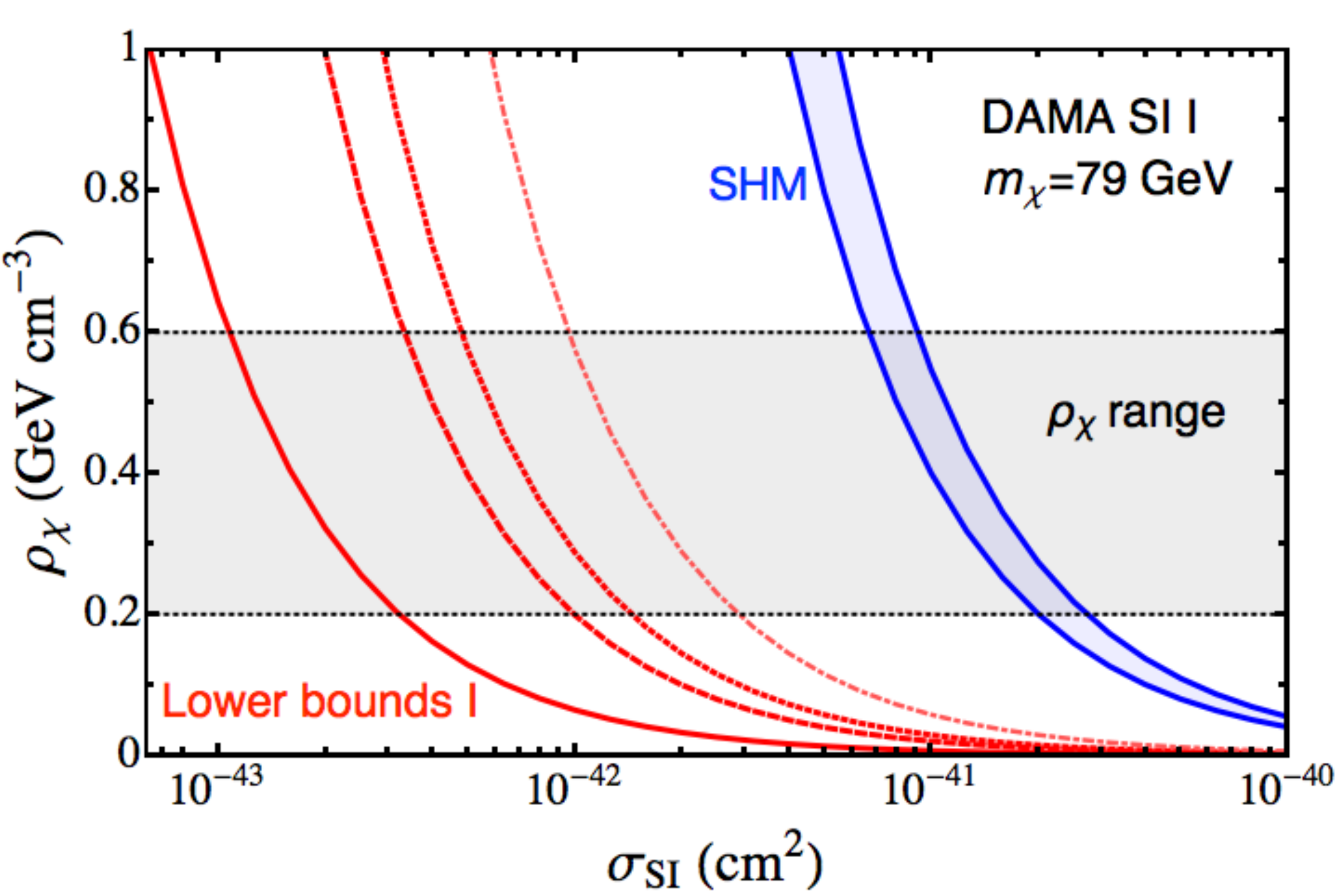}	\\
	\includegraphics[width=0.50\textwidth]{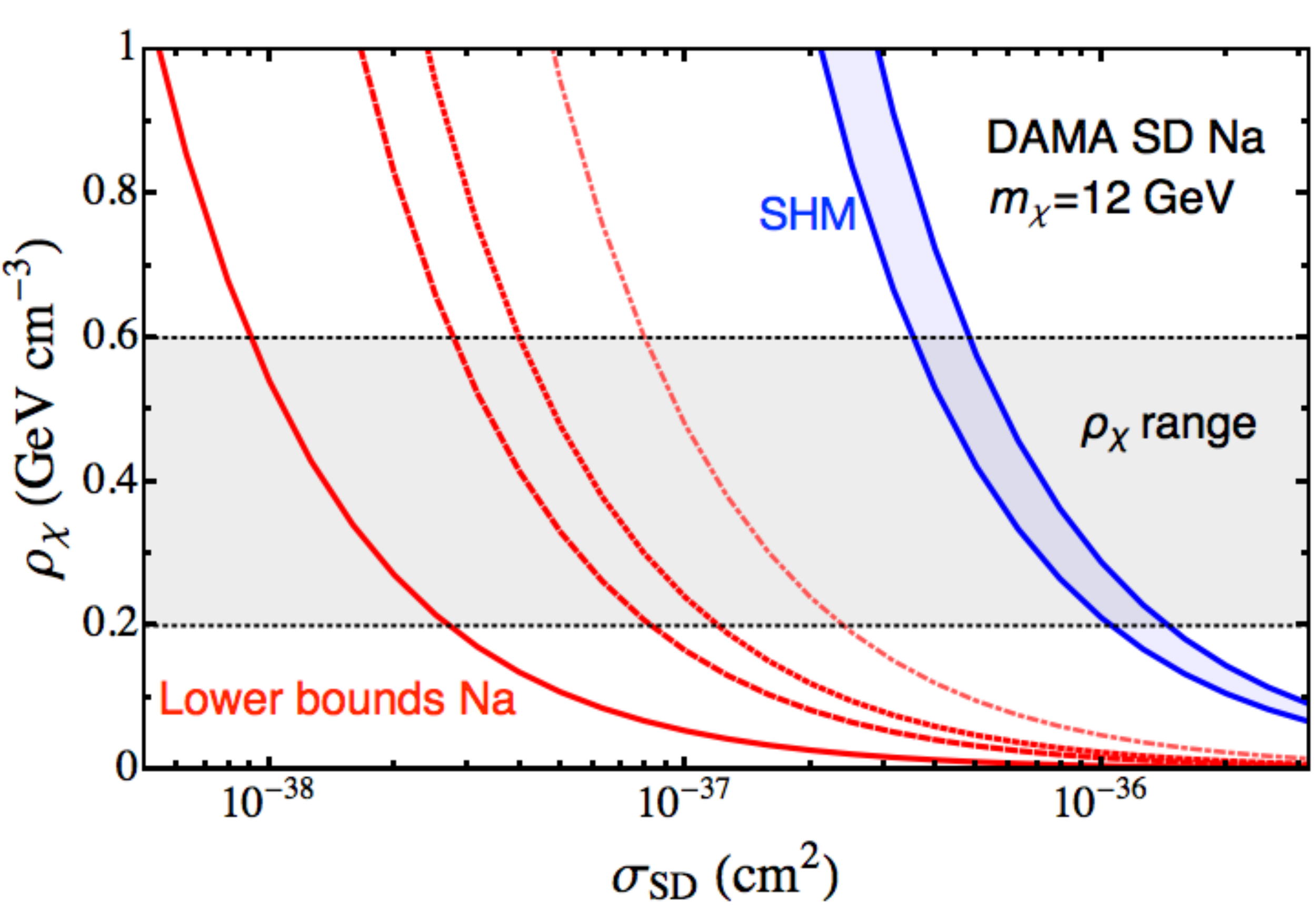}~~\includegraphics[width=0.50\textwidth]{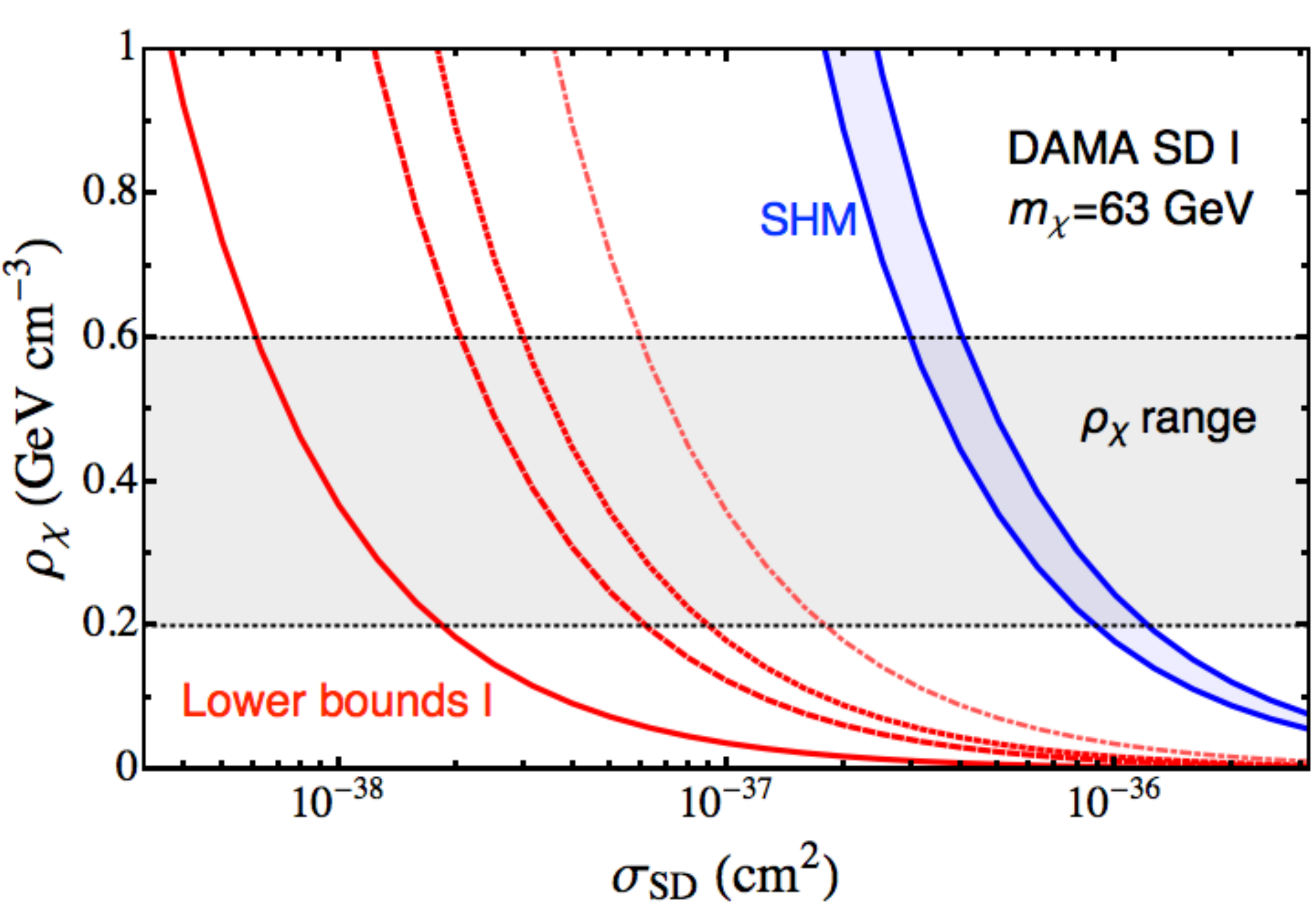}
	\caption{Lower bounds on the local energy density applied to DAMA modulation SI (top) and SD (bottom), and for scattering on Na (left) and I (right). The DM masses in each case are taken to be close to the best-fit points. We show the SHM prediction (solid blue), together with the bounds in red, from left to right: the ``Spectrum bound" of eq.~\eqref{eq:bound2mod} (solid red), the ``General bound" of eq.~\eqref{eq:final_bound} (dashed red) and the ``Symmetric bound" of eq.~\eqref{eq:final_boundb} (dotted red for $\sin \alpha=1$, dotted-dashed red for $\sin \alpha=0.5$. Notice that the latter is a factor of 2 stronger than the former.). All are shown at $90\%\,\rm CL$.} \label{DAMArho}
\end{figure}

This is shown in figure \ref{DAMArho}, where we plot the local energy density versus the scattering cross section. We show in blue the $90\%\rm \,CL$ SHM predictions and in red the $90\%\rm \,CL$ lower bounds, for SI (top) and SD (bottom), and for scattering on Na (left) and I (right). We plot from left to right the ``Spectrum bound" of eq.~\eqref{eq:bound2mod} (solid red), the ``General bound" of eq.~\eqref{eq:final_bound} (dashed red) and the ``Symmetric bound" of eq.~\eqref{eq:final_boundb} (dotted red for $\sin \alpha=1$, dotted-dashed red for $\sin \alpha=0.5$). As expected, the last three give much stronger constraints than the former.

For the very conservative ``General bound" (dashed red) we obtain that SI cross sections $\sigma_{\rm SI} \lesssim 8\cdot 10^{-42} \,(3\cdot 10^{-43})\, \rm cm^2$ are disfavoured for $m_\chi=12$ GeV, scattering on Na ($m_\chi=79$ GeV, scattering on I). For SD $\sigma_{\rm SD} \lesssim 3\cdot 10^{-38}\,(2\times 10^{-38}) \,\rm cm^2$,  for $m_\chi=12$ GeV, scattering on Na ($m_\chi=63$ GeV, scattering on I). A bit stronger constraints are obtained for the ``Symmetric bound", eq.~\eqref{eq:final_boundb} (dotted red), for $\sin \alpha=1$ (if the DM flow is aligned with the velocity of the Sun, i.e., $\sin \alpha=0.5$, the limits, shown with a dotted-dashed red curve, are a factor of 2 stronger than in this last case). Compared to the SHM, these are roughly between $\sim0.5$ and $\sim1.5$ orders of magnitude weaker.

Notice that these cross sections are excluded independently of the unknown velocity distribution. In addition, the limits are conservative in the sense that they still hold if the particle that gives rise to the DD signal constitutes only part of the total DM, since then the lower bound on the total local DM density can only be larger.

\section{Conclusions} 
\label{sec:conclusions}
In this paper we have derived new halo-independent lower bounds on the product of the DM--nucleon scattering cross section and the local DM density that are valid for annually-modulated signals, extending the work done in ref.~\cite{Blennow:2015gta} for constant rates. First, we have used that the amplitude of the annual modulation should be smaller than the rate. This allows to obtain the ``Events bound", eq.~\eqref{eq:bound1mod}, if only the number of events is measured, and the ``Spectrum bound", eq.~\eqref{eq:bound2mod}, if the spectrum is observed. Then we have derived new tests by making use of halo-independent bounds obtained in refs.~\cite{HerreroGarcia:2011aa, HerreroGarcia:2012fu} by doing an expansion of the rate on the Earth's velocity. These are the main results of the paper: the ``General bound", eq.~\eqref{eq:final_bound}, valid for general haloes, and the ``Symmetric bound", eq.~\eqref{eq:final_boundb}, which assumes some preferred direction for the DM velocity. These bounds can also be extended to multi-target detectors, see appendix~\ref{ap:multi}.

In order to illustrate their use, we have applied them to DAMA data. In combination with local energy measurements, we are able to exclude cross sections that are roughly between one and one and a half orders of magnitude smaller than the SHM ones, see figure~\ref{DAMArho}, but with the reward of being independent of the unknown velocity distribution. It is important to stress that these limits must be fulfilled by any particle physics model that gives rise to elastic SI/SD interactions in order for the DAMA signal to be consistent with DM.

In the spirit of being as conservative as possible, we would like to encourage the community to show, for positive DM direct detection signals, in addition to the typical SHM best-fit regions, the halo-independent lower bounds derived in this work, for instance the ``Symmetric bound" of eq.~\eqref{eq:final_boundb} (dotted red) for the cases where the phase is constant, or the ``General bound" of eq.~\eqref{eq:final_bound} (dashed red) when the phase varies, in a similar fashion as was done in figures~\ref{DAMA} and~\ref{DAMAmulti} for the DAMA signal.

\bigskip

{\bf Acknowledgements:} I am very grateful to Thomas Schwetz for fruitful discussions regarding not only this work but also our previous related projects. I would also like to thank Mattias Blennow and Stefan Vogl for useful discussions on our recent related project.
\appendix


\section{Derivation of the upper bounds on the modulation}
\label{ap:centraleq}

We derive here eq.~\eqref{eq:boundboth_gen} valid for the ``General halo". By applying eq.~\eqref{eq:bound2} in the ``General halo", eq.~\eqref{eq:bound_gen2}, we get:
\begin{equation} \label{eq:formula0}
\int_{v_{1}}^{v_{2}} \negthickspace \negthickspace dv A_\eta(v)
\leq  v_e\left(\frac{1}{v_1} -\frac{1}{v_1} \int_{v_1}^{v_2} dv \, \overline\eta(v)+
\int_{v_{1}}^{v_{2}} \negthickspace \negthickspace
dv \frac{1}{v^2}\,-\int_{v_{1}}^{v_{2}} \negthickspace \negthickspace
dv  \frac{1}{v^2}\,\int_{v}^{v_2} dv^\prime \, \overline\eta(v^\prime) \right) \,.
\end{equation}
Integrating the third term, and by parts the last term, we obtain for eq.~\eqref{eq:formula0}:
\begin{equation} \label{eq:formula1}
\int_{v_{1}}^{v_{2}} \negthickspace \negthickspace dv A_\eta(v)
\leq  v_e\left(\frac{2}{v_1} -\frac{1}{v_2} -\frac{2}{v_1} \int_{v_1}^{v_2} dv \, \overline\eta(v)+\,\int_{v_1}^{v_2} dv \, \frac{1}{v}\,\overline\eta(v) \right) \,.
\end{equation}
Now we can use that the last term of eq.~\eqref{eq:formula1} obeys
\begin{equation}  \label{eq:formula2}
\int_{v_{1}}^{v_{2}} \negthickspace \negthickspace dv\, \frac{1}{v}\,\overline\eta(v) \leq \frac{1}{v_1} \int_{v_{1}}^{v_{2}} \negthickspace \negthickspace dv\,\overline\eta(v)\,,
\end{equation}
and inserting eq.~\eqref{eq:formula2} back into eq.~\eqref{eq:formula1}, we get:
\begin{equation} \label{eq:formula3}
\int_{v_{1}}^{v_{2}} \negthickspace \negthickspace dv A_\eta(v)
\leq  v_e\left(\frac{2}{v_1} -\frac{1}{v_2} -\frac{1}{v_1} \int_{v_1}^{v_2} dv \, \overline\eta(v)\right) \,.
\end{equation}  
The last term of eq.~\eqref{eq:formula3} depends on $\overline{\eta}(v)$ and we can conservatively drop it. We finally obtain:
\begin{equation}
\int_{v_{1}}^{v_{2}} \negthickspace \negthickspace dv A_\eta(v) \leq  v_e\left(\frac{2}{v_1} -\frac{1}{v_2} \right) \,,
\label{eq:boundboth_genc}
\end{equation}
which is an upper bound on the modulation amplitude expressed solely in terms of velocities. In the end the result is the same as if we had used in eq.~\eqref{eq:bound_gen2} the weaker bound of~eq.~\eqref{eq:bound1} instead of the stronger one of eq.~\eqref{eq:bound2}. An analogous derivation can be performed for eq.~\eqref{eq:boundboth_spec}. 

Notice that strictly speaking all these inequalities should read ``less than" ($<$), instead of ``less or equal than" ($\leq$), because: 1) eq.~\eqref{eq:bound2} (and similarly eq.~\eqref{eq:bound1}) is very conservative, assuming just that $\overline\eta$ is a non-increasing function of velocity and, 2) eq.~\eqref{eq:bound_gen2}, and therefore eq.~\eqref{eq:boundboth_gen} (and similarly eq.~\eqref{eq:bound_spec2} and \eqref{eq:boundboth_spec} for symmetric haloes), are based on a first order expansion on the Earth's velocity. Moreover, in the last step we dropped a positive term.\footnote{Unless $v_1=v_2$, which is of course not an interesting case as then there is no range in velocity in which a modulation has been observed.} However, we decide to keep ``less or equal than" in all inequalities to make explicit that for very special (and probably unrealistic) haloes the inequalities could be (at least approximately) saturated, see ref.~\cite{HerreroGarcia:2011aa} for examples. In practice this means that any modulation signal that even saturates the bounds is very unlikely have a DM origin.

\section{Binned versions of the bounds}
\label{ap:bin}

In order to apply the bounds to real data, we need binned expressions, which necessarily involve some approximations. We will discuss the validity of the binning prescriptions here used at the end of the section. Let us define the bin average of a quantity $X(E)$ as 
\be
\langle X \rangle_i \equiv \frac{1}{\Delta E_i} \int_{E_{i1}}^{E_{i2}} dE\,X(E) \,,
\ee
where $E_{i1}$ and $E_{i2}$ are the boundaries of bin $i$ and $\Delta E_i =
E_{i2} - E_{i1}$.

We denote by $\mathcal{M}_i\equiv \langle \mathcal{M} \rangle_i$ the observed modulation in bin $i$, where $i=1,...,N$. For a fixed DM mass, $A_\eta$ can be obtained from the amplitude of the modulation in bin $i$, i.e., $\mathcal{M}_i$, by:
\be\label{eq:Atilde}
A_\eta(v_i) \equiv A_\eta^i = \frac{\mathcal{M}_i \,q}
{\mathcal{C}\,A^2 \langle F_A^2 \rangle_i f_A}  \,,
\ee
where $q$ is a possible quenching factor which we assume to be energy-independent (as is the case of DAMA), $v_i\equiv v_m(E_i)$ is the minimum velocity centered in the corresponding energy bin using eq.~\eqref{eq:R}, $\langle
F_A^2 \rangle_i$ is the nucleus form factor averaged over the bin
width and $f_A$ is the nucleus mass fraction if different elements or isotopes contribute to the rate. Similarly, $\overline{\eta}_i\equiv \overline\eta(v_i)$ can be expressed in terms of the constant rate $\mathcal{\overline R}_i$.

Now we have to perform a bin average of the bounds. This involves quantities like $\langle g(v)\, \mathcal{M} (E_R)\rangle_i$, which we replace by $\langle g(v)\rangle_i\, \mathcal{M}_i$, which is a good approximation whenever $g(v)$ does not vary drastically within a bin. The integral is approximated by a sum over bins. Using eq.~\eqref{eq:Atilde}, the ``Spectrum\, bound", eq.~\eqref{eq:bound2mod}, becomes:
\begin{equation}\label{eq:bound2modbin}
  \rho_\chi \sigma_{\rm SI} \ge \frac{2 m_\chi \mu^2_{\chi p}}{A^2} \left( \frac{q\,\langle v \rangle_1}{f_A\,\langle F_A^2 \rangle_1}\,\mathcal{M}_1 +    \sum_{i=1}^N \,\frac{q\,\Delta v_i}{f_A\,\langle F_A^2 \rangle_i}\,\mathcal{M}_i \right), 
\end{equation}
where we denote by $\Delta v_i$ the width in velocity space of bin $i$.

Similarly, for the ``General\, bound", eq.~\eqref{eq:final_bound}, we get:
\be \label{eq:final_boundbin} 
\rho_\chi \sigma_{\rm SI} \geq \frac{2 \,m_\chi\,\mu_{p}^2}{A^2}\,\frac{1}{v_e}\left(\left \langle \frac{2}{v} \right\rangle_1-\left \langle \frac{1}{v} \right\rangle_2\right)^{-1} \,\sum_{i=1}^N \,\frac{q\,\Delta v_i}{f_A\,\langle F_A^2 \rangle_i}\,\mathcal{M}_i\,.
\ee
And finally the ``Symmetric\, bound", eq.~\eqref{eq:final_boundb}, reads in its binned version:
\begin{equation} \label{eq:final_boundbbin} 
\rho_\chi \sigma_{\rm SI} \geq 
\frac{2 \,m_\chi\,\mu_{p}^2}{A^2}\,\Big\langle \frac{1}{v} \Big\rangle_1^{-1}\,\frac{1}{\sin\alpha\, v_e}\, \,  \sum_{i=1}^N \,\frac{q\,\Delta v_i}{f_A\,\langle
F_A^2 \rangle_i}\,\mathcal{M}_i \, .
\end{equation}
The inevitable inaccuracies involved in the binning procedure are small whenever the quantities within a bin do not show drastic changes. By using different prescriptions for the bounds, for instance $1/v_1, \,\text{vs} \,\,\langle 1/v \rangle_1,\,\text{vs}\, \,1/\langle v \rangle_1$, we have checked that inaccuracies due to the binning procedure are negligible. Notice also that the averaging of the velocity $\langle v \rangle$ can be performed analytically, as the relation with recoil energy $E_R$ is known, c.f., eq.~\eqref{eq:R}. In addition we need to extract information from the measured modulation amplitude in a particular bin (of size $0.5$ keVee for DAMA), which necessarily implies unfolding form factors and the detector response, which involves energy resolution, efficiencies and quenching factors. We have also checked that by including energy resolution in a similar fashion as in ref.~\cite{Kelso:2013gda} (see also refs.~\cite{Bernabei:2008yh, Bernabei:2012zzb}) the quantities involved in our bounds do not vary much, and therefore energy resolution is not expected to change the results significantly.

\section{Bounds for multi-target detectors}
\label{ap:multi}

With the exception of the ``Events bound", eq.~\eqref{eq:bound1mod}, in order to apply the rest of the bounds here derived to multi-target experiments one needs to assume that scattering is dominated by a particular nucleus. Let us now generalize the bounds to experiments with different $n$ elements, labeled by $A$. For a similar procedure for multi-target bounds see ref.~\cite{HerreroGarcia:2011aa}. In this case the modulation amplitude in an energy bin $i$ receives contributions from each element: $\mathcal{M}_i=\sum_{A=1}^n \mathcal{M}_i^A$. We use the notation $v_i^A\equiv v_m^A(E_i)$, see eq.~\eqref{eq:R}. Summing eq.~\eqref{eq:bound2} in its binned version for all the $n$ elements we get:
\begin{align}  
n &\geq\sum_{A=1}^n \overline \eta(v_1^A)\, \langle v^A\rangle_1+ \sum_{A=1}^n \sum_{i=1}^N \Delta v_i^A \, \overline\eta(v_i^A)\,\\
& =\sum_{A=1}^n\, \frac{q\,\langle v^A\rangle_1}{  \mathcal{C}\,A^2 \langle F_A^2 \rangle_1 f_A}\,\mathcal{\overline R}_1^A + \sum_{A=1}^n \,\sum_{i=1}^N\, \,\frac{q\, \Delta v_i^A}{\mathcal{C}\,A^2 \langle F_A^2 \rangle_i f_A}\,\mathcal{\overline R}_i^A \label{eq:naprox} \\
& \geq \frac{1}{\mathcal{C}}\,  \left(\min_A\left(\frac{q\,\langle v^A\rangle_1}{A^2 \langle F_A^2 \rangle_1 f_A}\right)\, \mathcal{\overline R}_1 +\,\sum_{i=1}^N\, \min_A\left(\frac{q\,\Delta v_i^A }{A^2 \langle F_A^2 \rangle_i f_A}\right)\,\mathcal{\overline R}_i   \right), \label{eq:naproxb}
\end{align}
where $\mathcal{\overline R}_i=\sum_{A=1}^n\mathcal{\overline R}_i^A$ is the total rate in bin $i$, and $\min_A(X_i)$ means that we need to take the minimum $X$ between all the nuclei present in the target, for each bin $i$. From the first to the second line we used the equivalent of eq.~\eqref{eq:Atilde} for $\overline{\eta}_i$ (in terms of constant rates), and from the second to the third line we changed the order of the sums and used the fact that $\mathcal{C}$ is detector independent, see eq.~\eqref{eq:C}. Using that constant rates are larger than modulations, $\mathcal{\overline R}_i\geq\mathcal{M}_i$, and the definition of $\mathcal{C}$, eq.~\eqref{eq:C}, we get:
\begin{equation}\label{eq:boundspect_multi}
  \rho_\chi \sigma_{\rm SI} \ge \frac{2 m_\chi \mu^2_{\chi p}}{n}\,
  \left( \min_A\left(\frac{q\,\langle v^A\rangle_1}{A^2 \langle F_A^2 \rangle_1 f_A}\right)\,\mathcal{M}_1 + 
        \,\sum_{i=1}^N\,\min_A\left(\frac{q\,\Delta v_i^A }{\,A^2 \langle F_A^2 \rangle_i f_A}\right)\,\mathcal{M}_i
  \right) \,,
\end{equation}
which is the ``Spectrum bound" valid for multi-target experiments.

Now we can do a similar thing for the bounds based on an expansion on the Earth's velocity, the ``General bound", eq.~\eqref{eq:final_bound}, and the ``Symmetric bound", eq.~\eqref{eq:final_boundb}.  Summing eq.~\eqref{eq:boundboth_gen} in its binned version for all the $n$ elements we get:
\begin{equation} 
\frac{1}{\mathcal{C}}\,\sum_{A=1}^n \sum_{i=1}^N\, \frac{q\, \Delta v_i^A }{\,A^2 \langle F_A^2 \rangle_i f_A}\,\mathcal{M}_i^A \leq v_e\,\left( \sum_{A=1}^n \,\left \langle \frac{2}{v^A}\right\rangle_1 \,-\sum_{A=1}^n\,\left\langle \frac{1}{v^A}\right\rangle_2 \right)\,.
\end{equation}
By changing the order of the sums, we finally obtain:
\begin{equation}\label{eq:boundgen_multi}
  \rho_\chi \sigma_{\rm SI} \ge \frac{2 m_\chi \mu^2_{\chi p}}{v_e}\,
\left( \sum_{A=1}^n \,\left \langle \frac{2}{v^A}\right\rangle_1 \,-\sum_{A=1}^n\,\left\langle \frac{1}{v^A}\right\rangle_2 \right)^{-1}\, \sum_{i=1}^N\,\min_A\left(\frac{q\, \Delta v_i^A}{\,A^2 \langle F_A^2 \rangle_i f_A}\right)\,\mathcal{M}_i\,  \,,
\end{equation}
which is the ``General bound", eq.~\eqref{eq:final_bound}, for multi-target experiments.
Doing the same procedure in eq.~\eqref{eq:boundboth_spec}, we get for the  ``Symmetric bound", eq.~\eqref{eq:final_boundb}:
\begin{equation}\label{eq:boundsym_multi}
  \rho_\chi \sigma_{\rm SI} \ge \frac{2 m_\chi \mu^2_{\chi p}}{\sin\alpha\,v_e}\,
\left( \sum_{A=1}^n \,\left \langle \frac{1}{v^A}\right\rangle_1\right)^{-1}\, \sum_{i=1}^N\,\min_A\left(\frac{q\, \Delta v_i^A}{\,A^2 \langle F_A^2 \rangle_i f_A}\right)\,\mathcal{M}_i\, \,,
\end{equation}
which is now valid for multi-target experiments. 

It is easy to check how these multi-target bounds reduce to the single-target ones for $n=1$, as it has to be. Notice however that the bounds here derived are weak due to the crude approximations needed in going from eq.~\eqref{eq:naprox} to eq.~\eqref{eq:naproxb}. For instance, for two isotopes of the same element, the multi-target bounds should be as strong as the mono-target ones, but due to the simplifications made they are clearly weaker by more than a factor of $2$.

\begin{figure}
	\centering
	\includegraphics[width=0.65\textwidth]{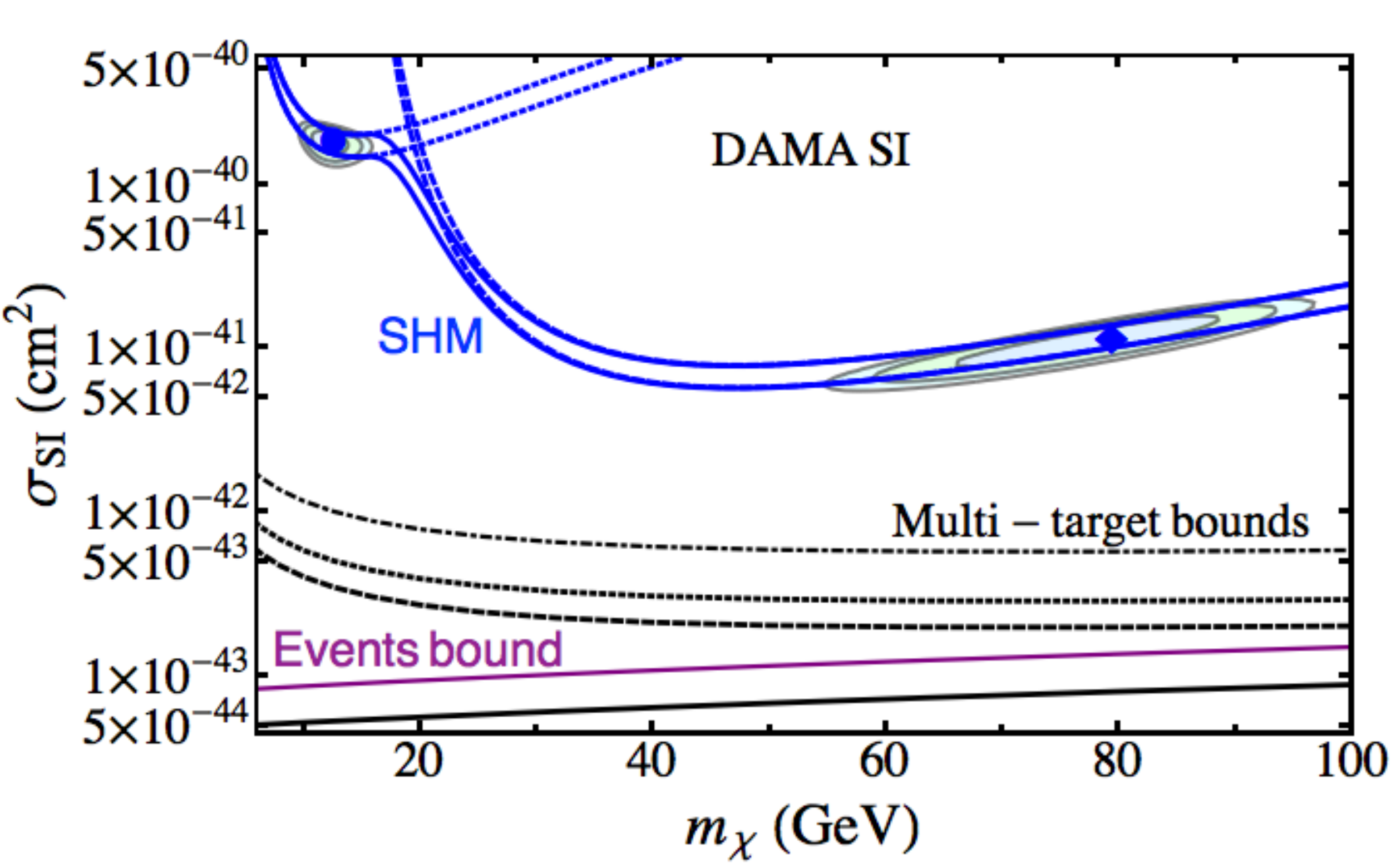}\\
	\includegraphics[width=0.65\textwidth]{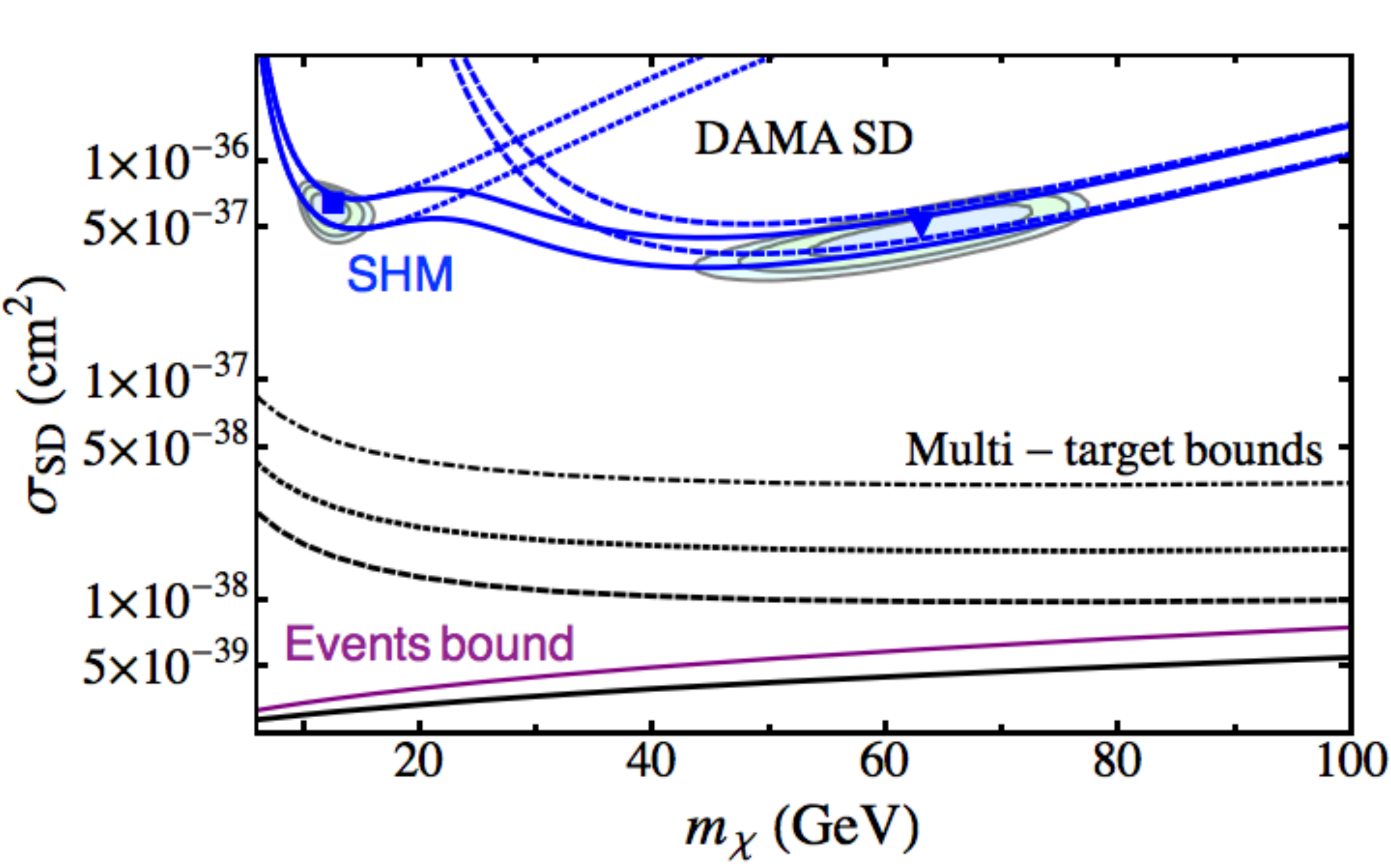}
	\caption{Bounds (in black) applied to DAMA modulation for SI (top) and SD (bottom), shown at $90\%\,\rm CL$. The $90\%\,\rm CL$ SHM range is plotted in blue (dotted for scattering on Na, dashed on I, and solid on both) together with the ``Spectrum bound" of eq.~\eqref{eq:boundspect_multi} (solid), the ``General bound" of eq.~\eqref{eq:boundgen_multi} (dashed) and the ``Symmetric bound" of eq.~\eqref{eq:boundsym_multi} (dotted for $\sin \alpha=1$, dotted-dashed for $\sin \alpha=0.5$), using both Na and I.  The ``Events bound" of eq.~\eqref{eq:bound1mod}, which is valid for multi-target, is plotted as solid purple. The SHM $\Delta\chi^2=2.3,5.99,9.21$ contours corresponding in two dimensions to a CL of $68.27, 95, 99 \%$ are shown in blue, green and light blue, respectively, together with the best-fit points, depicted with blue marks.} \label{DAMAmulti}
\end{figure}

In figure \ref{DAMAmulti} we show in black from bottom to top the ``Spectrum bound" of eq.~\eqref{eq:boundspect_multi} (solid), the ``General bound" of eq.~\eqref{eq:boundgen_multi} (dashed) and the ``Symmetric bound" of eq.~\eqref{eq:boundsym_multi} (dotted for $\sin \alpha=1$, dotted-dashed for $\sin \alpha=0.5$) for DAMA data for SI (top) and SD (bottom), using both sodium and iodine. The ``Events bound", eq.~\eqref{eq:bound1mod}, which is valid for multi-target, is plotted as solid purple. We also show the SHM prediction in blue (dotted for scattering on Na, dashed on I, and solid on both). The SHM $\Delta\chi^2=2.3,5.99,9.21$ contours (already shown in figure~\ref{DAMAbf}) corresponding in two dimensions to a CL of $68.27, 95, 99 \%$ are shown in blue, green and light blue, respectively, together with the best-fit values, depicted with blue marks. As expected, in all cases the multi-target bounds are weaker than the ones obtained for the single-target bounds, cf. figure~\ref{DAMA}. In particular, for SI interactions in the low DM mass region ($m_\chi\lesssim 20$ GeV) the difference between the multi-target bounds and the SHM is greater than two orders of magnitude, as the former ones are suppressed by the iodine $A^2$ factor with respect to the latter, which are dominated by scatterings on sodium. Notice that the ``Events bound", eq.~\eqref{eq:bound1mod} (solid purple), is stronger than the ``Spectrum bound" for multi-target detectors (solid black) due to the $1/n$ suppression of the latter ($n=2$ in this case), see eq.~\eqref{eq:boundspect_multi}.

\bibliographystyle{JHEP.bst}
\bibliography{mod}

\end{document}